\documentclass[sigconf]{acmart}
\AtBeginDocument{%
  \providecommand\BibTeX{{%
    \normalfont B\kern-0.5em{\scshape i\kern-0.25em b}\kern-0.8em\TeX}}}

\copyrightyear{2024} 
\acmYear{2024} 
\setcopyright{rightsretained} 
\acmConference[CHI '24]{Proceedings of the CHI Conference on Human Factors in Computing Systems}{May 11--16, 2024}{Honolulu, HI, USA}
\acmBooktitle{Proceedings of the CHI Conference on Human Factors in Computing Systems (CHI '24), May 11--16, 2024, Honolulu, HI, USA}
\acmDOI{10.1145/3613904.3642919}
\acmISBN{979-8-4007-0330-0/24/05}

\usepackage{hyperref}

\usepackage{xcolor}
\usepackage{soul}

\usepackage{balance}

\makeatletter
\AtBeginDocument{\let\hl\@firstofone}
\makeatother

\usepackage{appendix} 

\begin{document}

\title{The Effects of Generative AI on Design Fixation and Divergent Thinking}






\author{Samangi Wadinambiarachchi}
\affiliation{%
  \institution{The University of Melbourne}
  \city{Melbourne}
  \country{Australia}
  \postcode{3000}}
  \email{samangi.w@unimelb.edu.au}

\author{Ryan M. Kelly}
\affiliation{%
  \institution{RMIT University}
  \city{Melbourne}
  \country{Australia}
  \postcode{3000}}
  \email{ryan.kelly@rmit.edu.au}

\author{Saumya Pareek}
\affiliation{%
  \institution{The University of Melbourne}
  \city{Melbourne}
  \country{Australia}
\postcode{3000}}
  \email{spareek@student.unimelb.edu.au}

\author{Qiushi Zhou}
\affiliation{%
  \institution{The University of Melbourne}
  \city{Melbourne}
  \country{Australia}
\postcode{3000}}
  \email{qiushi.zhou@unimelb.edu.au}

\author{Eduardo Velloso}
\affiliation{%
  \institution{The University of Melbourne}
  \city{Melbourne}
  \country{Australia}
\postcode{3000}}
  \email{eduardo.velloso@unimelb.edu.au}

\renewcommand{\shortauthors}{Wadinambiarachchi et al.}

\begin{abstract}


Generative AI systems have been heralded as tools for augmenting human creativity and inspiring divergent thinking, though with little empirical evidence for these claims. This paper explores the effects of exposure to AI-generated images on measures of design fixation and divergent thinking in a visual ideation task. Through a between-participants experiment (N=60), we found that support from an AI image generator during ideation leads to higher fixation on an initial example. Participants who used AI produced fewer ideas, with less variety and lower originality compared to a baseline. Our qualitative analysis suggests that the effectiveness of co-ideation with AI rests on participants' chosen approach to prompt creation and on the strategies used by participants to generate ideas in response to the AI's suggestions. We discuss opportunities for designing generative AI systems for ideation support and incorporating these AI tools into ideation workflows.

\end{abstract}





\maketitle
\section{Introduction}

Consider a team of designers discussing ideas for environmentally friendly transport solutions for a city. One team member kicks off the discussion with a suggestion about electric buses. The rest of the team then spends an hour discussing variations on this idea, all involving electric vehicles, until an intern who arrived late asks ``have you considered bicycles?''. Until the intern's suggestion, the ideas were anchored on a salient characteristic of the first proposal---an electric motor.
The design literature dubs this phenomenon \textit{design fixation}---the ``blind adherence to a set of ideas or concepts limiting the output of conceptual design''~\cite[p.~1]{Jansson1991DesignFixation}. This is a common experience in any creative task, from art to engineering, and happens when exposure to one idea anchors and biases subsequent ideas, restricting exploration of the design space. Fixation happens both consciously and unconsciously, regardless of the level of experience of the practitioner~\cite{Jansson1991DesignFixation, Youmans2014DesignPrevention} and in all areas of creative work. The severe negative impact that design fixation has on the creative process makes it a key concern in design studies.

In the initial stages of the design process, it is common for designers to conduct precedence studies and create mood boards by compiling external stimuli as sources of inspiration to broaden their ideation space~\cite{Lucero2012FramingWork}. However, the exposure to previous solutions during this process can potentially be a source of design fixation. Previous studies have shown that exposure to examples of similar design solutions has mixed effects on creativity~\cite{Youmans2014DesignColors}. It tends to drive designers towards the example, narrowing the explored solution space~\cite{Jansson1991DesignFixation, Leahy2020DesignIdeas}. Further, variations in the modality~\cite{Viswanathan2016AFixation, Vasconcelos2016InspirationChallenges}, the fidelity ~\cite{Cheng2014ADesigners, Vasconcelos2016InspirationChallenges},  the quality~\cite{Vasconcelos2016InspirationChallenges}, the diversity and novelty of the exposed stimuli, the time of exposure, and its proximity to the design problem \cite{Vasconcelos2016InspirationChallenges} can vary the intensity of design fixation~\cite{Vasconcelos2017InspirationGeneration}.

Recent developments in generative artificial intelligence (GenAI) have been heralded as the harbinger of a new paradigm of creative work, often under the guise of augmenting human creativity~\cite{Eapen2023HowCreativity}.  Publicly available AI image generators such as DALL·E\footnote{\url{https://openai.com/dall-e-2}}, Artbreeder\footnote{\url{https://www.artbreeder.com}}, Stable Diffusion\footnote{\url{https://stablediffusionweb.com}}, and Midjourney\footnote{\url{https://www.midjourney.com}} have made it possible for designers to turn their thoughts into high-quality visuals quickly and at a low cost. The ability of these tools to generate ``original'' images based on user prompts potentially offers a rich source of inspiration. For example, Chiou et al.~\cite{Chiou2023DesigningGenerators} have shown that when used in co-ideation tasks, AI can open up a broader conceptual space quickly and effortlessly, promoting divergent thinking ~\cite{Chiou2023DesigningGenerators}. However, \textbf{there is still a lack of empirical evidence for the effect of generative AI as a source of inspiration during design tasks}. Though the specific outputs generated by these tools are novel, they are trained on existing work, blurring the lines between what is original and derivative. Further, designers could still be fixated during the ideation process despite any potential inspiration from AI.

In this paper, we aim to understand the effects of AI-generated imagery as a source of inspiration in an ideation task. We conducted a between-participants experiment in which designers took part in a visual ideation task that involved sketching ideas for a chatbot avatar. We manipulated participants' access to sources of inspiration: none, access to Google Image Search, or access to Midjourney (an image generation AI tool). Through our study, we sought answers to the following questions: 

\begin{itemize}
    \item RQ 1: How does the exposure to AI-generated images affect design fixation and divergent thinking during ideation, compared to using commonly used sources of inspiration and no inspiration support?
    \item RQ 2: How do different ways of interacting with AI image generators impact participants' effectiveness in an ideation task?  
\end{itemize}

We evaluated the effect of inspiration sources on participants' ideation output (the sketches). In doing so, we used four divergent thinking measures from prior literature (design fixation score, fluency, variety, and originality~\cite{Jansson1991DesignFixation, Shah2003MetricsEffectiveness, Youmans2014DesignPrevention, Vasconcelos2016FluencyExplanation}) to assess different facets of their creative output. We found that exposure to AI-generated images induced higher design fixation in participants than in other conditions. Moreover, fluency, variety, and originality were lower in the AI-supported group compared to the baseline condition. Through our qualitative analysis, we suggest that fixation arises when creating prompts and when ideating in response to AI images. In addition, we demonstrate that using AI can result in \textit{fixation displacement}, where the focus of fixation shifts from an exemplar onto the AI's outputs.


Our study provides an empirical contribution to the AI-powered creativity support literature by illustrating how AI-generated images influence design fixation and divergent thinking measures. It further elaborates on AI's role in providing inspiration during visual design tasks. Further, we demonstrate the importance of focusing on factors that might induce design fixation while acquiring inspiration from AI tools and propose potential strategies and directions to explore in mitigating design fixation.

\section{Related Work}

Our research builds on studies of design fixation and on the role AI can play in supporting design ideation. 

\subsection{Design Fixation }

Among the factors that hinder designers' creativity, \textit{``design fixation''} is one of the most well-studied phenomena in creativity and design research. It is identified as the \hl{unconscious adherence to a set of pre-known ideas or knowledge that restricts the ideation space}~\cite{Jansson1991DesignFixation, Youmans2014DesignPrevention}. When a person experiences design fixation, they tend to adhere to pre-conceived ideas and concepts, limiting exploration of the design space during ideation~\cite{Linsey2010AFaculty, Purcell1996DesignFixation, Jansson1991DesignFixation}. 
Design fixation narrows designers' ability to explore the creative space between abstract ideas and potential solutions~\cite{Vasconcelos2016InspirationChallenges, Jansson1991DesignFixation}. Previous findings show that this is reflected heavily in their design outcomes and restrains designers from maximising their creative potential, resulting in unoriginal outputs~\cite{Jansson1991DesignFixation}. 

Design fixation has been studied extensively across different fields~\cite{Vasconcelos2018IdeaExperiments}, including  cognitive science~\cite{Cao2021UtilizingGeneration}, design~\cite{Bellows2013AnResearch, Kim2013ToIntentionality}, education~\cite{Howard2013OvercomingMethods}, mechanical engineering~\cite{Youmans2011TheFixation,Viswanathan2012DesignCost}, and psychology~\cite{Smith2013ShiftingFixation, Bellows2012TheFixation}. These studies have collectively shown that design fixation is more likely to occur when designers are exposed to example solutions for design tasks~\cite{Jansson1991DesignFixation}. It has also been demonstrated that the modality, degree of abstraction of the inspiration (i.e. the fidelity), and the designer's level of expertise~\cite{Alipour2018AFactors} can affect fixation intensity when exposed to external stimuli.

Fixation has typically been studied through quantitative experimental approaches in which participants are asked to solve a design problem, either with or without an example (external stimuli)~\cite{Vasconcelos2016InspirationChallenges}. For instance, Jansson and Smith's classic design fixation work~\cite{Jansson1991DesignFixation}  reported four experiments. These experiments divided participants into two groups: a treatment group (fixation group), who were given a design problem along with an example solution, and a control group, who were given the same problem with no examples to work from. They hypothesised that showing an example design would restrict the ideas of the treatment group because it would make the participants fixate on the given example. Jansson and Smith~\cite{Jansson1991DesignFixation} found that even though both groups produced a similar number of designs, ideas in the fixation groups were more similar to the example. In a subsequent experiment, the researchers found that the flexibility and originality of the designs were limited in the fixation group and concluded that creative performance may be inhibited when an example induces design fixation.  Since then, several studies have been conducted replicating or amending the method and examples~\cite{Vasconcelos2016InspirationChallenges, Leahy2020DesignIdeas}.

\hl{When looking at design fixation, it is important to distinguish different types of fixation effects}~\cite{Crilly2017WhereDesign}.\hl{ Youmans and Arciszewski}~\cite{Youmans2014DesignPrevention}\hl{ identify three such effects. The first is \textit{unconscious adherence}}~\cite{Youmans2014DesignPrevention} \hl{to past designs without realizing. An example of this is copying the features of an example (even if the features are inappropriate to the task)}~\cite{Jansson1991DesignFixation, Crilly2017WhereDesign}.\hl{The second is \textit{conscious blocking}}~\cite{Youmans2014DesignPrevention}\hl{, where new ideas are actively but perhaps momentarily dismissed. In this situation, a designer is aware of alternative creative paths but chooses to disregard them, perhaps due to a commitment to a current project's direction or a bias towards familiar solutions. The third type of fixation effect is \textit{intentional resistance}, a deliberate decision against exploring new concepts. For instance, design companies engaged in research and development often prefer to explore solutions that fall within their well-established expertise, a tendency known as \textit{local search bias}}~\cite{Rosenkopf2001BeyondIndustry, Crilly2017WhereDesign}.

Apart from trying to understand its causes, researchers have explored various strategies to overcome design fixation~\cite{Vasconcelos2016InspirationChallenges}. Such strategies include incorporating physical prototyping in the ideation activities~\cite{Viswanathan2012DesignCost}, triggering frequent reminders for participants to consider all available options in a timely manner during an ideation task~\cite{Luchins1942MechanizationEinstellung., Youmans2014DesignPrevention}, utilising design thinking and lateral thinking methods \cite{Belski2015ApplicationExperts} such as de Bono’s six thinking hats \cite{deBono2008SixHats, Andersson2012DesignUnseen}, having short breaks or ``incubation periods'' during the task~\cite{Youmans2011DesignFixation, Smith2011AFixation}, using computer-aided design and intelligent agents \cite{Davis2019CreativeCreativity, Hoggenmueller2023CreativeExplorations}, and incorporating design by analogy \cite{Casakin1999ExpertiseEducation}. 

Even though there is a large body of work on design fixation exploring the effects of external stimuli on creative tasks~\cite{Alipour2018AFactors}, studies centred around design fixation are limited within the field of HCI. Among these few studies, HCI researchers have started to examine the potential of using AI image generators as tools for supporting creativity~\cite{Lewis2023AIxArtist:Block,Hoggenmueller2023CreativeExplorations}. Thus, in this study, we adapt experimental methods from mechanical engineering and design research,  \hl{where design fixation is framed as \textit{unconscious adherence} and is measured by the degree to which participants directly copy features from an example stimulus}. We aim to understand the influence of AI image generators on design fixation and divergent thinking, adding new empirical evidence to the HCI literature.

\subsection{The Emergent Role of AI in Creativity Support}

Since the early 1990s, designers have envisioned a future with intelligent design and creative aids~\cite{Gero1994DesignAids}. With recent advances in Generative AI, this vision is becoming a reality. Generative AI systems can create new, plausible media~\cite{Remy2020EvaluatingResearch}  to aid individuals in creative tasks~\cite{Joon2021TheTools}. Generative AI models are trained on large data sets and can enable people to generate content such as images, text, audio, or video quickly and easily~\cite{Koch2020ImageSense:Partnerships}. Currently, Generative AI tools enable users to create diverse artefacts by providing instructions in natural language called ``prompts''. Generative AI systems can also synthesise diverse concepts and generate unpredictable ideas. In the case of AI image generators --- the focus of our study --- the output comes from the latent space of a deep learning model, arising from an iterated diffusion process that involves the model arranging pixels into a composition that makes sense to humans~\cite{Verheijden2023CollaborativeAI}.  Because of process randomness, different results can be obtained based on the same prompt, with entirely new images each time. This differs from conventional image searches, where the search is performed by entering a query into a database to retrieve images that the search engine considers relevant. Another difference is that whereas long and specific queries might be too restrictive for an image search engine, they can benefit AI image generators.

Previous works have explored the roles that generative AI can play in the creative process~\cite{Hwang2022TooProcesses}. For instance, AI can generate content entirely by itself with instructions from the user, or it can act as a creativity support tool, augmenting the user's creativity~\cite{Mazzone2019ArtIntelligence}. AI text generators can be used as a tool to define specific problems to solve and promote convergent and divergent thinking~\cite{Wingstrom2022RedefiningArtists} and have the potential to be used as a co-creative assistant for a designer~\cite{Davis2019CreativeCreativity,Shin2023IntegratingIdeation}. Professionals in creative industries claim that AI could be a promising tool to gather inspiration~\cite{AudiMediaCenter2022ReinventingMediaCenter}.

With the growing interest in AI, HCI researchers have also started to explore ways of using AI as a creativity support tool~\cite{Chung2022ArtisticTools, Joon2021TheTools}. Among these explorations, a growing stream of literature focuses on using generative AI to access inspiration and mitigate design fixation. Researchers speculate that generative AI will become a potential solution for inspiring designers~\cite{Lamiroy2022Lamuse:Inspiration, Sbai2019DesIGN:Networks, Singh2019CameraInspiration}  due to the ability of AI generators to create abstract and diverse stimuli~\cite{Karimi2020CreativeCo-creativityb}.

One of the early examples in HCI for incorporating AI to mitigate design fixation was the \textit{Creative Sketching Partner (CSP)}~\cite{Davis2019CreativeCreativity, Karimi2020CreativeCo-creativityb}, an AI-based creative assistant that generates inspiration for creative tasks. Through multiple studies, Davis et al. ~\cite{Davis2019CreativeCreativity} suggest that the CSP helped participants in ideation and in overcoming design fixation.  Hoggenmueller et al. have also explored how generative text-to-image tools can support overcoming design fixation experienced in the field of Human-Robot Interaction~\cite{Hoggenmueller2023CreativeExplorations}. They conducted a first-person design exploration and reflection using ``CreativeAI Postcards'' inspired by Lupi and Posavec's ``Dear Data book'' method to ideate and visualize robotic artefacts. They noted that AI-generated images have the potential to inspire new robot aesthetics and functionality and also claimed that the designer's AI-co-creativity can help to eliminate biases and expand limited imagination. In a different case, Lewis~\cite{Lewis2023AIxArtist:Block} reflects that a digital assistance tool like ``ChatGPT'' helped her by acting as an art teacher and providing instructions. Lewis points out that it is challenging to distinguish between inspiration and copying when utilizing generative AI and reflects on concerns such as \textit{``transparency of attribution''}, \textit{``ethical considerations''}, and the clarity of the \textit{``creation process''}. Rafner et al.~\cite{Rafner2023PictureProblem-Solving} conducted an in-the-wild study to examine the effects of AI-assisted image generation on creative problem-solving tasks, aiming to investigate the effects of generative AI on problem identification and problem construction. They developed a human-AI co-creative technology that combines a GAN and stable diffusion model to support AI-assisted image generation. They found that this intervention enabled participants to facilitate idea expansion and prompt engineering, suggesting that AI can ``aid users in generating new ideas and
refining their initial problem representations''~\cite{Rafner2023PictureProblem-Solving}. 

As the domain of AI-powered creativity support is still in its infancy, the available literature provides only a nascent understanding of the effect of AI on creativity and design fixation. Our work extends the literature by using established techniques from design fixation research to better understand how AI image generators affect design fixation during a visual design task.


\section{Method}

We conducted a between-participants experiment to understand how AI-generated imagery affects designers' divergent thinking during visual ideation after being exposed to an \hl{example design}. We compared this scenario to the use of online image search and to no inspiration support. The independent variable was the Inspiration Stimulus: none \textsc{(Baseline)}, Google Image Search \textsc{(Image search)}, or Generative AI \textsc{(GenAI)}. The dependent variables were the \textsc{Design Fixation} score (the number of features in each sketch in common with the example), \textsc{Fluency} (the number of sketches produced), \textsc{Variety} (the number of different types of sketches produced), and \textsc{Originality} (how infrequently other participants devised the same type of sketch). We conducted the experiment in a controlled laboratory setting following a mixed-method approach. All participants gave informed written consent to participate after reading a plain language statement describing the procedure. The study received ethics approval from our university.

\begin{figure*}[t]
    \centering
    \includegraphics[width=0.675\textwidth]{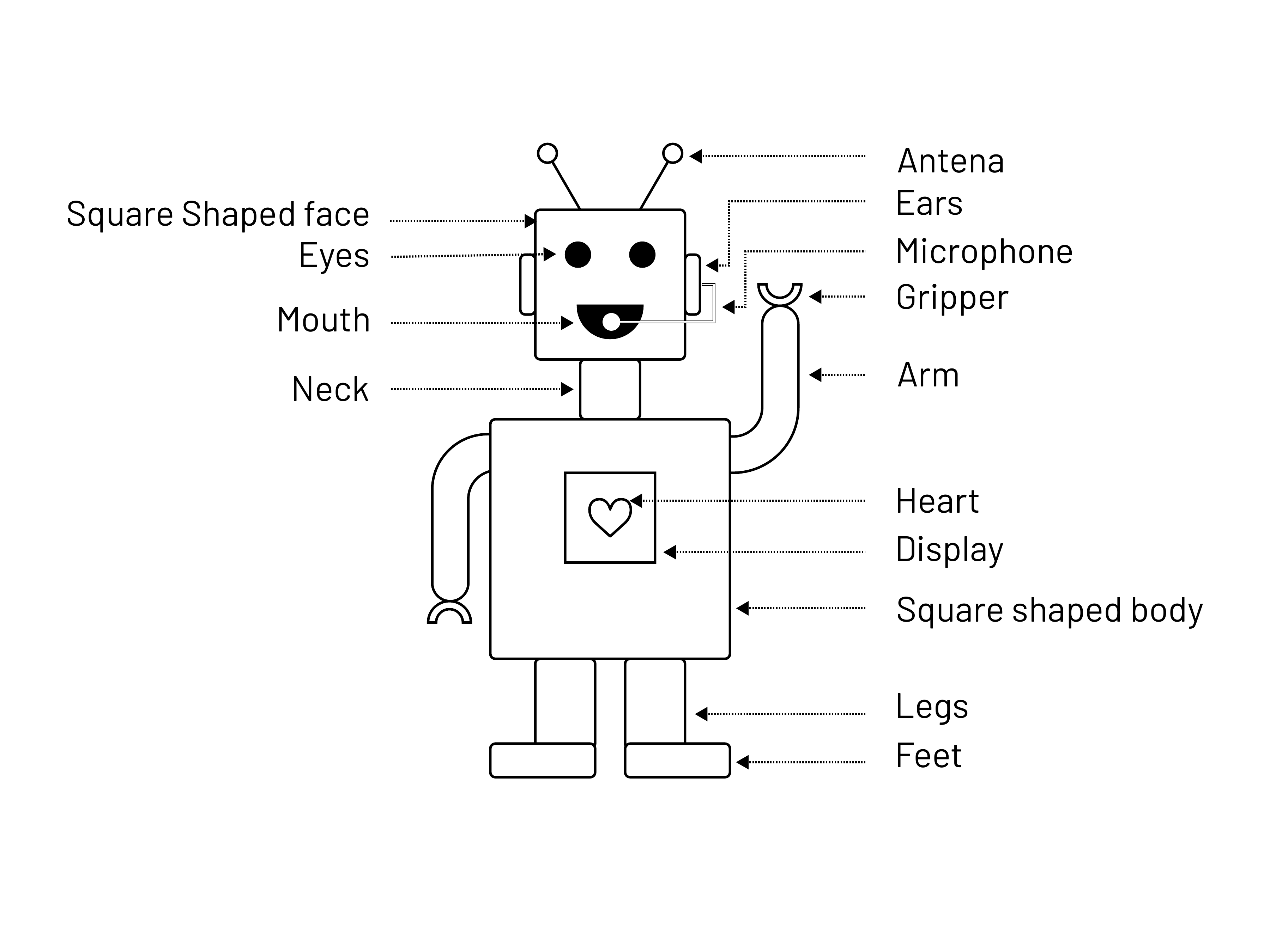}
    \vspace{-1.05em}
    \caption{The example with the 14 salient features we monitored. \textit{Note: The example was given to the participants without the callouts.}}
    \label{fig: example-given-with-the-brief}
    \vspace{-1.05em}
\end{figure*}

\subsection{Study Design and Materials}
\label{sec:materials}

The experimental task consisted of a visual ideation activity in which participants were asked to devise as many ideas as possible for a new chatbot avatar by sketching them on paper. \hl{The written design brief given to participants was:}
\begin{quote}
   \hl{ \textit{``Your task is to design a character we plan to use as an avatar for a chatbot. This chatbot is kind, loving, caring, and intelligent. It can assist you in solving your problems and is always there for you to talk to whenever you need to. So, imagine that you are conversing with this chatbot in real life and then come up with as many sketches as possible. Remember, you can annotate the sketch if you need to explain more about your design. And please always number each sketch you draw in the order you come up with them.''} }
\end{quote}

\hl{This written design brief included an example of an avatar with the figure caption "Example chatbot avatar (for reference only)". The example avatar is shown in Figure} \ref{fig: example-given-with-the-brief}.

\hl{Further, we provided verbal instructions for the participants, asking them to produce as many different ideas as they could during the experiment. For participants in the Image search and Gen AI conditions, we additionally informed them that they could use the digital tool (either Google Image Search or Midjourney, depending on the condition) to gather inspiration for their work. The full study protocol can be found in supplementary material.}

Similar to previous work~\cite{Jansson1991DesignFixation}, we started the task by showing participants an example avatar to induce design fixation. We drew inspiration from Ward's creature invention task~\cite{Kozbelt2007UnderstandingPredictors, Ward1994StructuredGeneration}, which asked participants to imagine and create animals that lived on a different planet. \hl{The authors of this paper created the example chatbot avatar after several design iterations.} We created the avatar so that it had 14 salient features, which we used to quantitatively assess design fixation (see Figure \ref{fig: example-given-with-the-brief}). We considered the presence of these features in participants' ideas to be evidence of design fixation, following standard practice in the literature~\cite{Jansson1991DesignFixation}. In the experimental task, participants were given 20 minutes to sketch their ideas for addressing the brief. \hl{We chose this time limit because it is the median time given to participants in previous design fixation studies }\cite{Vasconcelos2016InspirationChallenges} \hl{and because we aimed to cap each experimental session at one hour to avoid fatigue.} We provided participants with pencils, pens, felt pens, and coloured pencils, along with blank A4 sheets to sketch their ideas. A timer was placed outside their peripheral view for them to keep track of time.


The experiment included a single between-participants independent variable---the \textsc{Inspiration Stimulus} available during the task---with three levels:
\begin{itemize}
  \item \textsc{Baseline}: no inspiration support.
  \item \textsc{Image Search}: Participants had access to Google Images\footnote{\url{https://images.google.com/}} during the task, accessed through a web browser in incognito mode to avoid the browser history influencing results.
  \item \textsc{GenAI}: Participants had access to the paid version of Midjourney V4, an AI image generation tool, through a private Discord server running the Midjourney bot (which was required to enter prompts and view outputs from the model). Midjourney V4 was the default model when our study was conducted (May 2023)\footnote{\url{https://www.midjourney.com}}. Participants interacted with Midjourney through textual prompts that the model used to generate sets of four images per prompt.
\end{itemize}

\begin{figure*}[t]
\centering
    \includegraphics[width=0.8\textwidth]{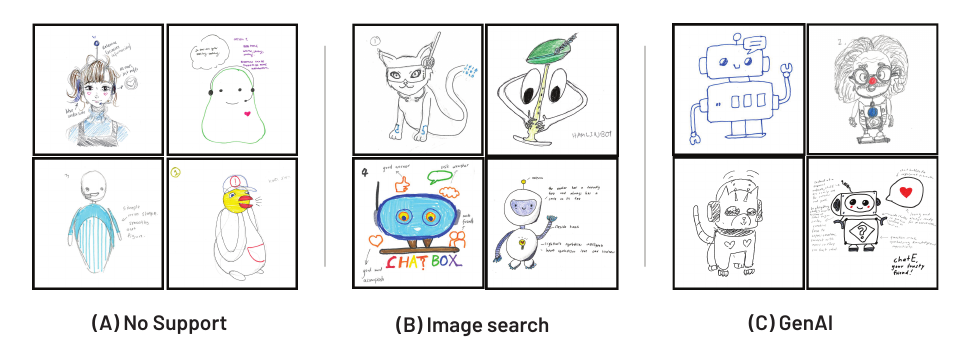}
    \caption{\hl{Examples of sketches created by participants in each experimental condition. (A) No support condition, (B) Image search condition, (C) GenAI condition}}
    \label{fig:sketchesByCond}
    \vspace{-1.05em}
\end{figure*}

We assessed participants' creative output using four standard measures from the design fixation literature: design fixation, fluency, variety, and originality, which we describe as follows:

\textsc{Design fixation} \hl{is the unintentional conformity towards existing ideas or concepts that limits exploration of the ideation space}~\cite{Jansson1991DesignFixation, Youmans2014DesignPrevention}. \hl{Researchers use the degree of copying as a method to quantify design fixation}~\cite{Jansson1991DesignFixation,Moreno2016OvercomingFindings}. Therefore, we operationalise design fixation as an objective property of each sketch based on the presence or absence of features available in the example. Following the approach used in design fixation literature \cite{Moreno2016OvercomingFindings}, two raters blind to the experiment's aims counted the presence of features from the \hl{example avatar} in the sketches created by the participants. We validated the ratings by computing the inter-rater reliability and computed the \hl{design fixation score (DFS)} as follows:

    {\small
    \begin{equation}
    \begin{split}
    \text{Design fixation score} = \frac{\text{Number of features repeated from the example}}{\text{Number of fixating features in the example }}\\
    \end{split}
    \end{equation}
    }

\textsc{Fluency} refers to the number of ideas produced by the participants ~\cite{Guilford1956TheIntellect.,Vasconcelos2016FluencyExplanation}. We operationalise it by counting the number of sketches produced by each participant within the available time (20 minutes).

    {\small
    \begin{equation}
    \begin{split}
    \text{Fluency} = \text{Number of sketches produced by the participant}\\
    \end{split}
    \end{equation}
    }

\textsc{Variety} measures the coverage of the solution space explored during the idea-generation process~\cite{Shah2003MetricsEffectiveness}. It aims to capture the extent of the design space covered during ideation.  If the majority of ideas are similar, it indicates less variety.  To compute variety, we assigned a numerical identifier to all the sketches (N=277), imported them into a Miro\footnote{\url{miro.com}} (an online collaborative whiteboard),  and displayed them in randomised order. Two raters (blind to the conditions) iteratively and inductively grouped similar sketches into mutually exclusive clusters. This activity considered several factors: appearance, embodiment, appendages, shape, and accessories. The process resulted in 83 clusters.
Each participant received a \textsc{Variety} score based on the number of clusters their sketches were classified into. We subtract 1 from the number of clusters so that if all of a participant's sketches belong to the same cluster, their score is 0, and if they have sketches in every cluster, their score is 1.
    
    {\small
    \begin{equation}
    \begin{split}
    \text{Variety} = \frac{\text{Number of clusters that a participant's sketches belong to - 1}}{\text{Number of clusters - 1}}\\
    \end{split}
    \end{equation}
    }

\textsc{Originality} (also called Novelty~\cite{Fiorineschi2023UsesLiterature, Shah2003MetricsEffectiveness}) refers to the uniqueness of a particular sketch within the total pool of sketches made by participants~\cite{Jansson1991DesignFixation, Guilford1956TheIntellect.}. It measures how unusual and unexpected a given idea is. Intuitively, the more people have the same idea, the less original it is. We computed an idea's originality by counting the number of other participants who had an idea belonging to the same cluster, dividing it by the total number of other participants, and computing its complement to 1 (to normalise the value between 0 and 1). In other words, it is the proportion of \textit{other} participants who did \textit{not} have the same idea. This score is 0 when every participant had an idea in the same cluster and 1 if only a single participant had an idea in that cluster.

    {\small
    \begin{equation}
    \begin{split}
    \text{Originality} = 1-\frac{\text{Number of other participants with ideas in the cluster}}{\text{Number of other participants}}\\
    \end{split}
    \end{equation}
    }

\begin{figure*}[t]
\centering
    \includegraphics[width=0.8\textwidth]{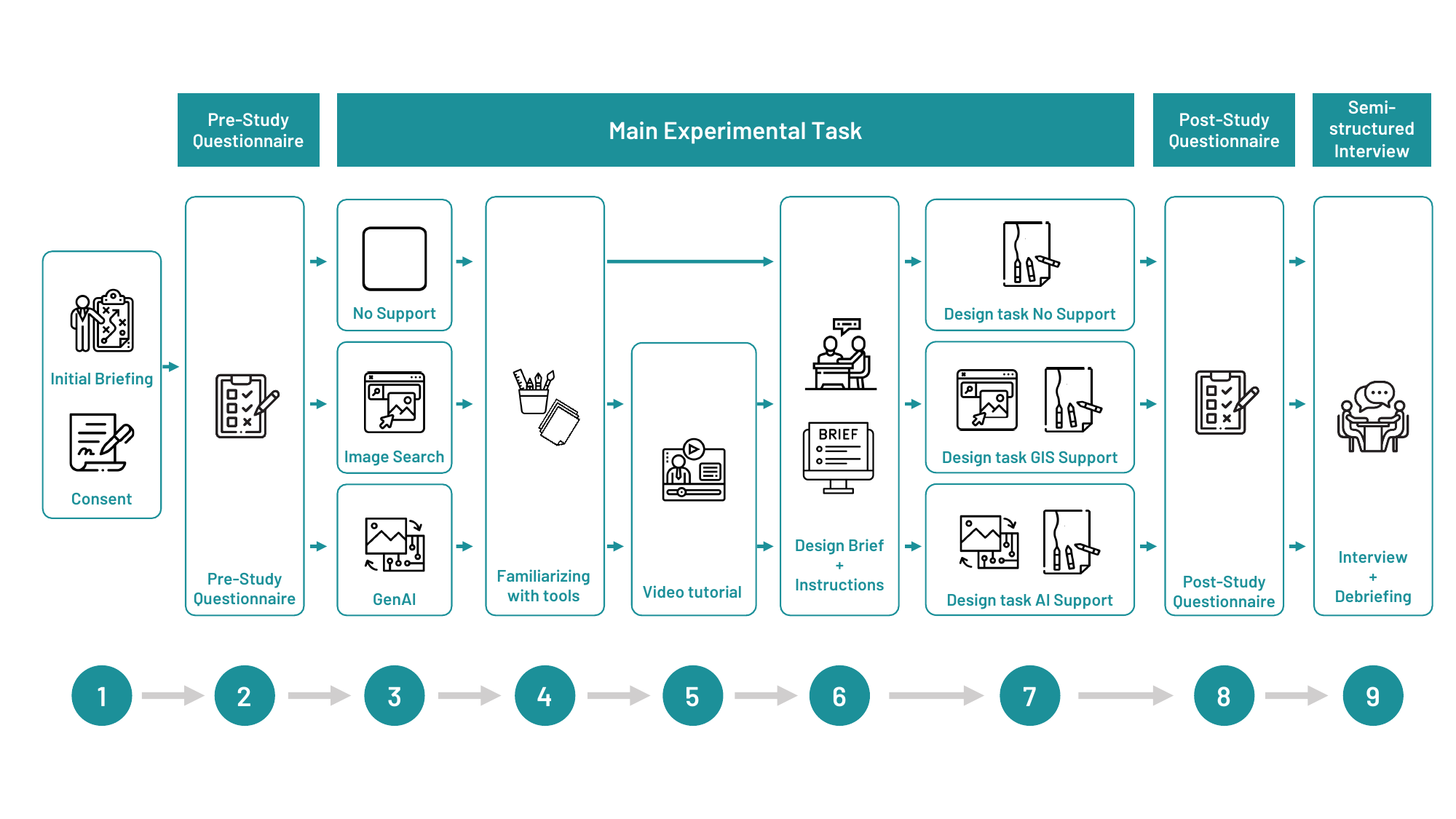}
    \caption{\hl{The overall experiment flow \textbf{1}: Initial briefing and participant consent, \textbf{2:} Pre-study questionnaire, \textbf{3-7}: Main experimental task, \textbf{8}: Post-study questionnaire, \textbf{9}: Semi-structured interview and debriefing}}
    \label{fig: method}
\end{figure*}

\subsection{Participants}

We recruited 60 participants through digital student notice boards, mailing lists of university student clubs, and word of mouth.  Participants expressed their interest through a digital signup form.\hl{ Participants self-described their prior experience in visual design (measured in years/months). We did not specify this experience should only be professional design experience}. We screened participants based on our eligibility criteria and invited those who were 18 years or older and had experience in visual design via email. Further, to avoid dependent relationships, we ensured that none of the participants had a direct connection with the primary researcher running the study. Participants had a mean age of 25.8 years (18--49, SD = 5.4). They included undergraduate, master's and PhD students from diverse domains such as arts, business, computer science \& IT, design, engineering, and science.  Each condition had an equal number of participants and was gender-balanced, with 10 women and 10 men per condition (gender was self-described by participants).

\subsection{Procedure}

Participants booked a time to participate individually based on their availability. The study was carried out in a quiet research laboratory. Upon arrival, participants read a plain language statement describing the study and consented to participate (Figure~\ref{fig: method}-1). 

The experiment had four stages: pre-study questionnaire, main experimental task, post-study questionnaire, and semi-structured interview. Each session lasted 45--60 min in total. In the \textbf{pre-study questionnaire}, we collected participants' basic demographic information, their experience with similar design tasks (measured in years/months), and their familiarity with AI image generators (a yes/no question, and participants were asked to list any systems they had used if they answered yes).
(Figure~\ref{fig: method}-2). The main objective of this questionnaire was to understand and control for any variables that might confound the results. After completing the questionnaire, the participants were randomly assigned to one of the three conditions and were assigned a unique ID generated by the computer (3 random digits) (Figure~\ref{fig: method}-3).

In the \textbf{main experimental task} (Figure~\ref{fig: method}, steps 3-7), participants in all conditions received the same design brief, which asked them to design an avatar for a chatbot in 20 minutes, as described in Section~\ref{sec:materials}. We started by allowing participants to familiarise themselves with the available materials. Then, the participants assigned to the image search and AI-supported groups received an introduction to the tool they would use during the design task (Figure \ref{fig: method}-5). These tools were available for them to use on an Apple MacBook M1 Pro laptop. The tool introduction included a video tutorial created by the research team. This video tutorial explained how to use the tool. After the video tutorial, we allowed participants to ask questions and clarify any doubts.

We provided task instructions to participants both verbally and as a written brief. The written brief included an example of a chatbot avatar, which served as a stimulus to induce design fixation (Figure \ref{fig: method}-6). Participants were given 20 minutes to complete the design task (Figure \ref{fig: method}-7). We limited the design task to 20 minutes to minimise the possibility of fatigue and because previous studies considered it an ideal duration for maintaining focus for producing ideas with both quality and quantity~\cite{Vasconcelos2016InspirationChallenges, Youmans2011TheFixation}. Once participants indicated they were ready to start, the researcher started the screen recording with participants' consent (in \textsc{Image Search} and \textsc{GenAI} conditions), switched on the timer and left them alone to work in the room, allowing them to work independently. 

After the design task, the researcher entered the room and asked the participant to fill in the \textbf{post-study questionnaire} (Figure \ref{fig: method}-8).  As the post-study questionnaire, we administered the NASA-TLX~\cite{Hart1988DevelopmentResearch} to ensure that all conditions induced an equivalent workload. \hl{To analyze the NASA-TLX, we used a one-way ANOVA; the effect of the independent variable "condition" on the NASA-TLX overall score was not statistically significant (F(2, 57) = 1, p = 0.37). Therefore, we did not conduct post-hoc tests.} 

Then, the researcher conducted a \textbf{semi-structured interview}. Each semi-structured interview lasted 15--20 min. Through the semi-structured interview, we aimed to get insights into the participant's background and their past experience in creating logos and avatars. We also probed for possible feelings of design fixation during the experiment and how it was affected by their previous knowledge, experience and process. In addition, we asked questions to understand how the stimuli (or lack thereof) affected their ideation process. To conclude the study, we debriefed the participants about the purpose of the research. We thanked each participant with a \$20 gift voucher.

\begin{figure*}[t]
\centering
    \includegraphics[width=1\textwidth]{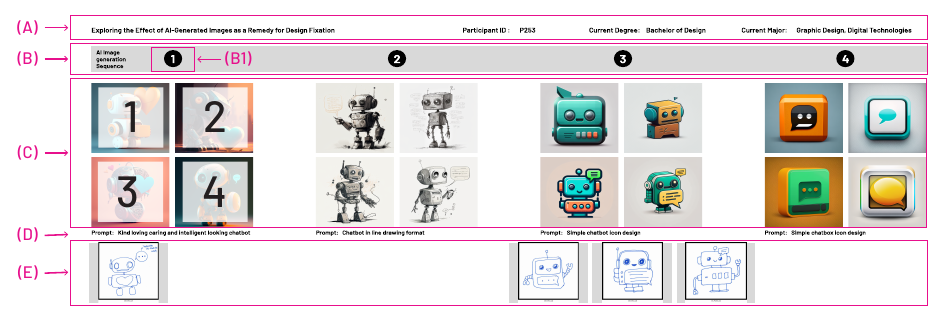}
    \caption{An example of visual sequence board\textbf{ (A)}: Participant information and meta data, \textbf{(B)}: AI image generation sequence, \textbf{(B1)}: Image generation number, \textbf{(C)}: AI-generated images in the order 1-2-3-4, \textbf{(D)}: Prompt used for each generation, \textbf{(E)}: Participant sketch sequence.}
    \label{fig:sequence_boards}
\end{figure*}

\subsection{Data Preparation}

We scanned all the sketches participants created and assigned them a unique identifier. Two independent evaluators rated the sketches to compute the design fixation score, variety, and originality measures. These evaluators were researchers from the human-computer interaction domain with experience in teaching and evaluating design. 

We extracted all prompts and images from the Midjourney gallery where the logs were saved (not visible to the participants), compiled the sketches and arranged them in the sequence in which they were created as a visual sequence board. Underneath the AI-generated images, we added the sketches of the participants (Figure \ref{fig:sequence_boards}).
\vspace{-0.65em}
\subsection{Data Analysis}

We used a mixed-method approach for our analysis. For quantitative analysis of design fixation and divergent thinking, we built Bayesian statistical models to quantify relationships between our dependent and independent variables (see \autoref{sec:stats}). We employ Bayesian statistical methods to analyze our results, opting for this approach due to its added flexibility, capability to quantify uncertainty, better handling of small samples, and greater potential for future extensibility. For a comprehensive rationale advocating the use of Bayesian methods over traditional frequentist statistics in the field of Human-Computer Interaction (HCI), see Kay et al. \cite{kay2016bayes}. Readers who may not be familiar with these methods can find a beginner-friendly introduction in McElreath~\cite{mcelreath2020statistical} and can see examples of their practical application in HCI in Schmettow~\cite{schmettow2021new}. In this manner, we shift the focus away from p-values and dichotomous significance testing, directing our discussion towards causal modelling and parameter estimation.

For qualitative analysis of participants' interview data, we used Braun and Clarke's 6-phase approach to reflexive thematic analysis  \cite{Terry2021EssentialsAnalysis, Braun2022ThematicGuide}. The analysis was inductive, i.e. data-driven, based on transcripts of the interviews. Each phase of the analysis was progressed using NVivo12\footnote{\url{https://lumivero.com/products/nvivo}} for coding procedures, theme development and naming. The analysis aimed to understand potential causes of design fixation during the experiment and participants' approaches to creating sketches in each condition. In this paper's findings, we use interview quotes to illustrate participants' approaches to prompt creation and their stated approaches to ideation based on AI images. This enables us to probe plausible explanations for observed differences between experimental conditions and explore why particular kinds of sketches were created in response to AI-generated images.
\vspace{1em}

\section{Results}

\subsection{Statistical Analysis}
\label{sec:stats}

\begin{figure}[h!]
\centering
    \includegraphics[width=0.4\textwidth]{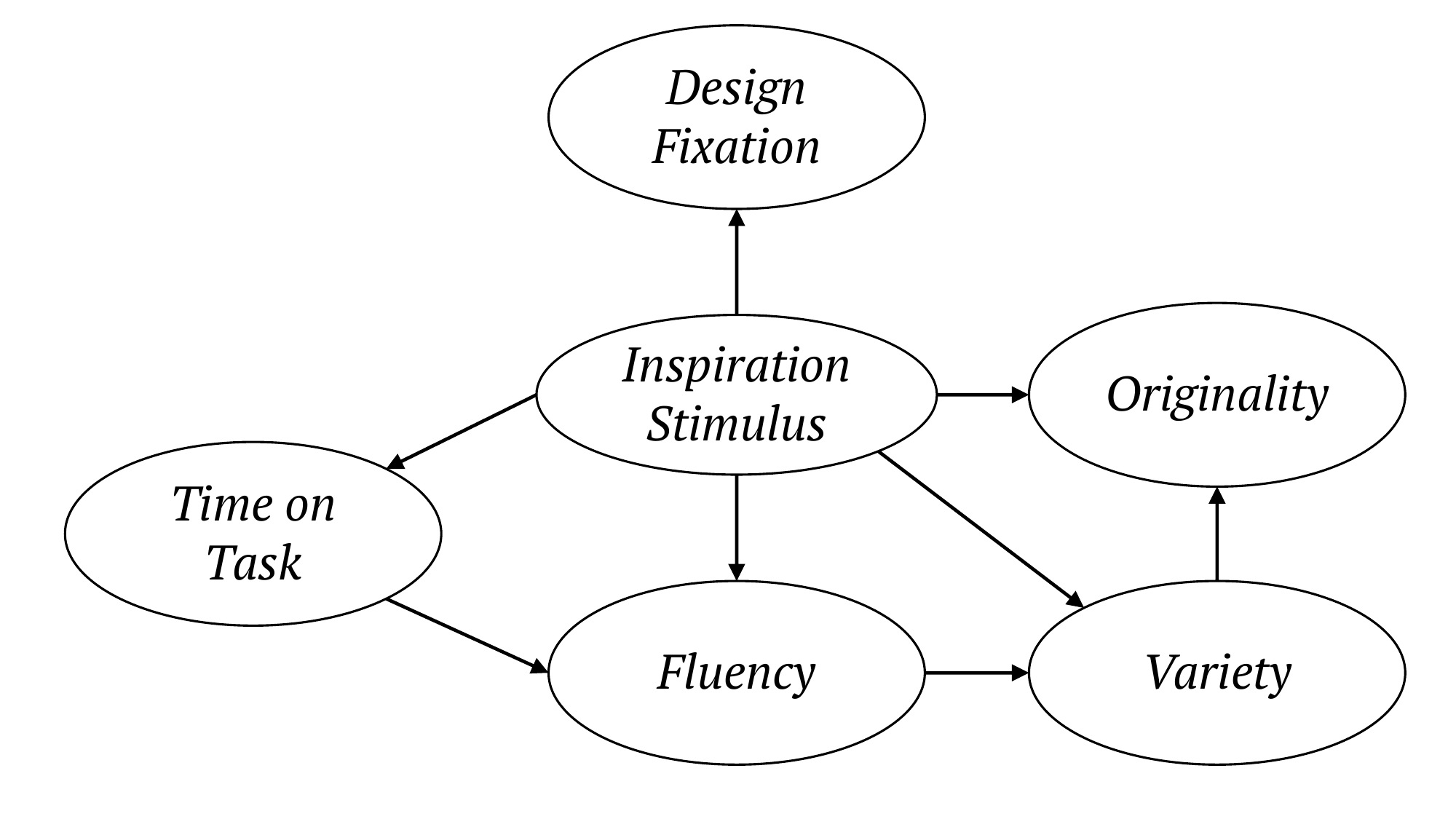}
    \caption{\hl{Theorised causal directed acyclic graph.}}
    \label{fig:dag}
\end{figure}

We summarise our theoretical claims as a directed acyclical graph (DAG) in Figure \ref{fig:dag}. We argue that the \textsc{Inspiration Stimulus} affects users' \textsc{Design fixation}, \textsc{Fluency}, \textsc{Variety}, and \textsc{Originality}.  The choice of inspiration stimulus affects how much time is spent on the sketching task (as opposed to seeking inspiration), which, in turn, affects the number of sketches produced (fluency). Higher fluency is also likely to lead to higher variety---as producing more sketches also increases the likelihood that they will cover more ground during ideation. A greater variety of sketches, in turn, will likely lead to more original ideas.

We used the \texttt{brms} package~\cite{burker2017brms} to fit our models. This package facilitates the implementation of Bayesian multilevel models in R, leveraging the Stan probabilistic programming language~\cite{carpenter2017stan}. To ensure the reliability of our Bayesian Markov Chain Monte Carlo (MCMC) sampling process, we assessed convergence and stability through two metrics: R-hat, which ideally should be less than 1.01~\cite{vehtari2021rhat}, and the Effective Sample Size (ESS), which should ideally exceed 1000~\cite{burker2017brms}. All of our model estimates met these criteria. We built our models based on the original count data in the direct measurements but report normalised values as described in Section \ref{sec:materials} in our plots for easier comparisons with future work. 

In our reporting of model results, we present the posterior means of parameter estimates, their corresponding standard deviations, and the boundaries of the 89\% compatibility interval, often referred to as the credible interval. The choice of an 89\% compatibility interval aligns with the recommendation by McElreath~\cite{mcelreath2020statistical} to mitigate potential confusion with the frequentist 95\% confidence interval, as the two intervals have distinct interpretations. The compatibility interval specifies the range of values within which there is an 89\% probability that the true value lies. We report hypothesis test results using Bayes Factors, which compares the likelihood of the observed data under the proposed model over the null. We interpret these values following Wagenmakers et al.~\cite{wagenmakers2011psychologists}, considering values above one as supporting a given hypothesis, values under 3 offering anecdotal evidence; under 10, substantial evidence; under 30, strong evidence; under 100, very strong evidence; and above 100, extreme evidence. We note that p-values are not used in Bayesian statistics, and no claims about ``statistical significance'' should be derived from our results.
\vspace{-0.75em}

\subsection{Design Fixation}
To model \textsc{Design Fixation}, we consider the number of salient features in participants' sketches also found in the \hl{example avatar} provided at the beginning of the experiment. 
We model this data as a binomial distribution with N = 14 (the maximum number of features) and a probit link. We use weakly informative, regularising priors for the model parameters (drawn from a normal distribution with mean zero and standard deviation of 2). We model the random effects of participants and images as being drawn from a normal distribution with mean zero and standard deviation computed from the data through partial pooling.

\begin{table}[h]
\centering
\caption{Summary of the binomial model for design fixation: \texttt{DFS|trials(14) $\sim$  Stimulus + (1|Participant ID) + (1|Image ID)}. We provide the posterior means of parameter estimates (Est.), posterior standard deviations of these estimates (SD), and the bounds of their 89\% compatibility interval. We note that this is not the same as the frequentist confidence interval but a percentile of the posterior distribution. All parameter estimates converged with an ESS well above 1000 and an R-hat of 1.00.}
\small{
\begin{tabular}{lll}
\toprule
\textbf{Parameter} & \textbf{Est. (SD)} & \textbf{89\% CI}\\
\midrule
Intercept                               & -.71 (.10)& [-.86, .56]\\
Image Search                            & .27 (0.14)& [.05, .48]\\
GenAI                                   & .32 (0.13) & [.11, .54]\\
\bottomrule
\end{tabular}
}
\label{tab:dfs_model}
\end{table}

\begin{figure}[h]
\centering
    \includegraphics[width=0.45\textwidth]{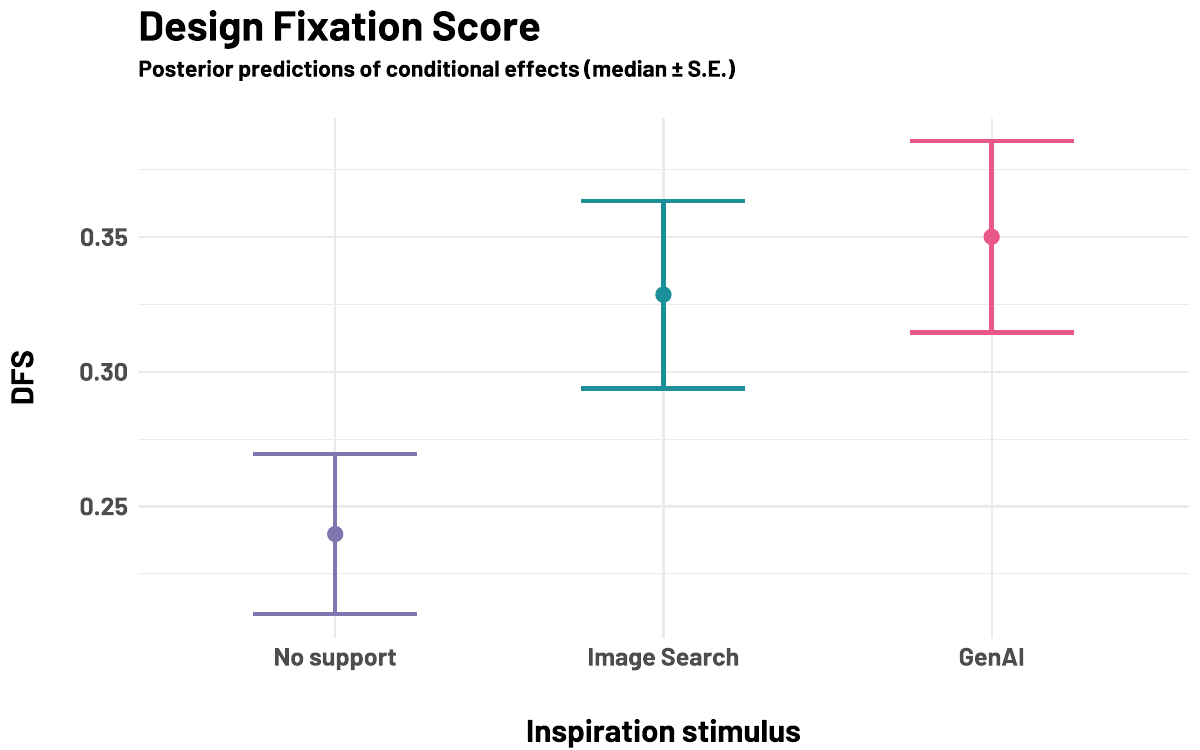}
    \caption{Model posterior predictions for \textsc{Design Fixation} scores. Error bars represent the standard error of the estimates. Scores correspond to the percentage of salient features in the example found in participants' sketches (higher is worse).}
    \label{fig:dfs_plot}
\end{figure}

The model suggests that the effect of \textsc{GenAI} has a 100\% probability of leading to higher \textsc{Design Fixation} (mean = .32, 89\% CI [.11, .55]), and a Bayes Factor of 124 suggests extreme support for the hypothesis that inspiration from \textsc{GenAI} leads to higher \textsc{Design Fixation} than the baseline. The effect of \textsc{Image Search} on \textsc{Design Fixation} was also detrimental, but less so (mean = .27, 89\% CI [.05, .49]), with a 98\% probability of this effect leading to higher \textsc{Design Fixation}. A Bayes Factor of 42.48 suggests very strong evidence for the hypothesis of a higher Design Fixation than the \textsc{No support} baseline. In summary, our model suggests that, \textbf{on average, both stimuli led to more features in common with the \hl{example avatar}, and \textsc{GenAI} led to even more design fixation than \textsc{Image Search}}.

\subsection{Fluency}
To model \textsc{Fluency}, we consider the number of sketches produced by each participant.  
\hl{Our causal model considers two effects of the stimulus on \textsc{Fluency}: a direct effect and an effect mediated by \textsc{Time on task (ToT)}.
We model these effects through two models, with and without \textsc{Offset(log(ToT))} as a covariate. This approach was taken to account for varying time on task as participants of both GenAI and Image search utilised different times to sketch.} 
In both cases, we model the expected value for each response as a negative binomial distribution with a log link. This models the sketch count based on the mean and the shape parameter, both of which depend on the inspiration stimulus. We opted for this model instead of a Poisson model due to its ability to model overdispersion. We use weakly informative, regularising priors for the model parameters (drawn from a normal distribution with mean zero and standard deviation of 2).

\begin{table}[h]
\centering
\caption{{\hl{Summary of the negative binomial models for \textsc{Fluency}---Direct effects: \texttt{Fluency $\sim$ + offset(log(Time on Task))} and Total Effects: \texttt{Fluency $\sim$  Stimulus}}. \hl{We provide the posterior means of parameter estimates (Est.), posterior standard deviations of these estimates (SD), and the bounds of their 89\% compatibility interval. We note that this is not the same as the frequentist confidence interval but a percentile of the posterior distribution. All parameter estimates converged with an ESS well above 1000 and an R-hat of 1.00.}}}
\small{
\begin{tabular}{lllll}
\toprule
& \multicolumn{2}{c}{\textbf{Direct Effects}} & \multicolumn{2}{c}{\textbf{Total Effects}}\\
\textbf{Parameter} & \textbf{Est.(SD)} & \textbf{89\% CI}  & \textbf{Est.(SD)} & \textbf{89\% CI} \\
\midrule
Intercept              &  -1.48 (.19)& [-1.79, -1.17]& 1.50 (.19) & [1.20, 1.80]\\
Image Search           & -.22 (.25)& [-.63, .17]& -.49 (.25)& [-.88, -.10]\\
GenAI                  & .10 (.31)& [-.39, .58]& -.21 (.30)& [-.66, .26]\\
Intercept$_{shape}$    & .75 (.45)& [.04, 1.47]&  .77 (.46)& [.04, 1.49]\\
Image Search$_{shape}$ & 1.58 (1.15)& [-.01, 3.61]& 1.66 (1.14)&[.06, 3.74]\\
GenAI$_{shape}$        & -.51 (.61)& [-1.49, .47]& -.40 (.65)&[-1.41, .66]\\
 \bottomrule
\end{tabular}
}
\label{tab:fluency_model}
\end{table}

\begin{figure}[h]
\centering
    \includegraphics[width=0.425\textwidth]{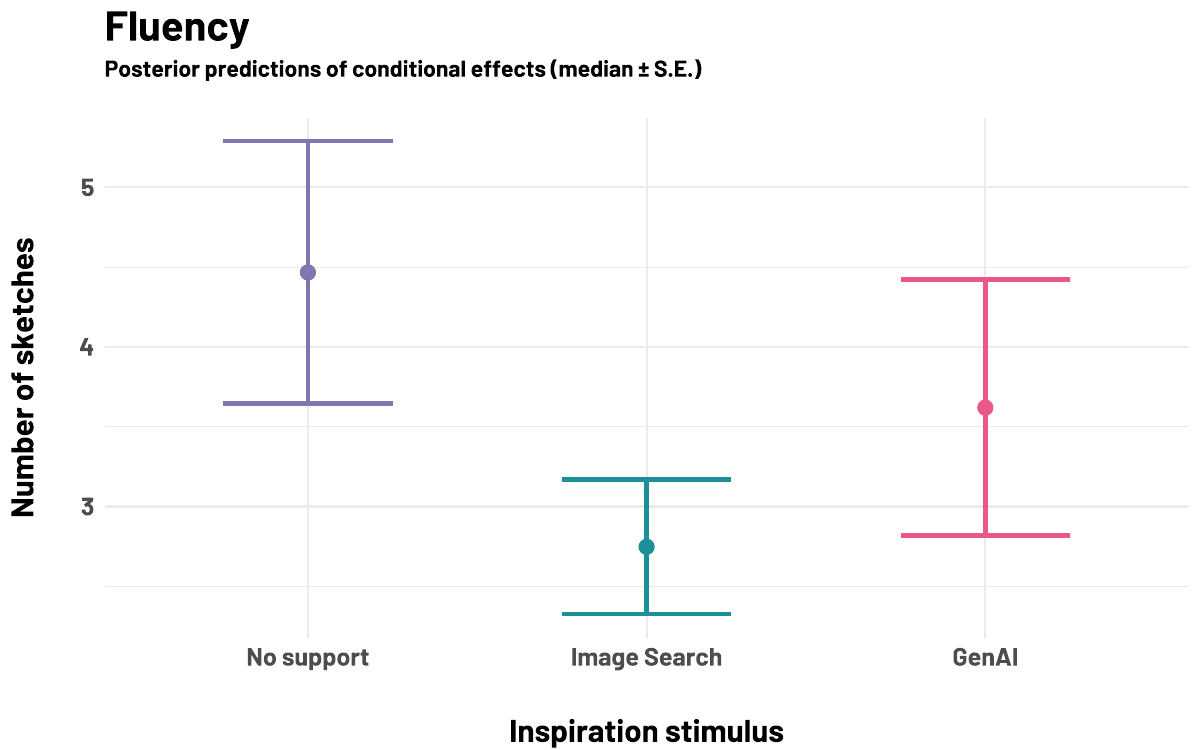}
    \caption{Model posterior predictions for \textsc{Fluency} (number of sketches generated by each participant). Error bars represent the standard error of the estimates.}
    \label{fig:fluency_plot}
    \vspace{-1em}
\end{figure}

\hl{In the total effects model, both \textsc{Image Search} and \textsc{GenAI} demonstrate detrimental effects on \textsc{Fluency}. Specifically, for \textsc{GenAI}, there is a notable negative impact on Fluency with an estimate of -.21 (89\% CI [-.7, .28]) and a Bayes factor of 3.20, suggesting a 76\% posterior probability of a negative effect. This is a substantial indication of its negative total effect on Fluency. The effect of \textsc{Image Search} is more pronounced (mean = -.49, 89\% CI [-.9, -.10]) with a Bayes factor of 46.62, indicating a 98\% posterior probability of a negative effect, strongly supporting its detrimental influence on Fluency.

In the direct effects model, which accounts for the time-on-task, the impact of \textsc{Image Search} on \textsc{Fluency} is minimal (mean = -0.22, 89\% CI [-0.64, 0.19]) with a Bayes factor of 4.20, indicating an 81\% posterior probability of a negative effect. In contrast, \textsc{GenAI} shows a relatively neutral direct effect on \textsc{Fluency} (mean = .10, 89\% CI [-.41, .60]) with a Bayes factor of .61, implying only a 38\% posterior probability of a negative effect.

In summary, \textbf{neither \textsc{Image Search} nor \textsc{GenAI} enhanced Fluency compared to the baseline, with both generally resulting in lower Fluency}. The effect of \textsc{Image Search} on Fluency is minimal when controlling for total output time, indicating a less direct impact. However, \textsc{GenAI} does not exhibit a considerable direct negative effect on \textsc{Fluency}, highlighting its influence is not strongly dependent on the total time available for sketching.}


\subsection{Variety}
To model \textsc{Variety}, we consider the number of clusters a participant's sketches belong to minus one to account for the fact that variety only begins with the second sketch. Our causal model considers two effects of the stimulus on \textsc{Variety}: a direct effect and an effect mediated by \textsc{Fluency}.
We model these effects through two models, with and without \textsc{Fluency} as a covariate.
In both cases, the expected value for each response is based on a negative binomial model with a log link. We use weakly informative, regularising priors for the model parameters (drawn from a normal distribution with mean zero and standard deviation of 2 for coefficients and a gamma distribution with parameters set to 0.01 for the shape).

\begin{table}[h]
\caption{Summary of the negative binomial models for \textsc{Variety}---Direct effects: \texttt{Variety $\sim$  Stimulus + Fluency} and Total Effects: \texttt{Variety $\sim$  Stimulus}. We provide the posterior means of parameter estimates (Est.), posterior standard deviations of these estimates (SD), and the bounds of their 89\% compatibility interval. We note that this is not the same as the frequentist confidence interval but a percentile of the posterior distribution. All parameter estimates converged with an ESS well above 1000 and an R-hat of 1.00.}
\small{
\begin{tabular}{lllll}
\toprule
& \multicolumn{2}{c}{\textbf{Direct Effects}} & \multicolumn{2}{c}{\textbf{Total Effects}}\\
\textbf{Parameter} & \textbf{Est. (SD)} & \textbf{89\% CI}  & \textbf{Est. (SD)} & \textbf{89\% CI} \\
\midrule
Intercept                           & .00 (0.23) & [-.38, .36]& 1.01 (.19)& [.71, 1.30]\\
Image Search                        & .02 (0.25) & [-.38, .42]& -.39 (.28)& [-.84, -.06]\\
GenAI                               & -.15 (0.24) & [-.53, .23]& -.29 (-.28)& [-.74, .17]\\
Fluency                             & .14 (0.02) & [.11, .18] &  & \\
\bottomrule
\end{tabular}
}
\label{tab:variety_model}
\vspace{-1.5em}
\end{table}

\begin{figure}[h]
\centering
    \includegraphics[width=0.425\textwidth]{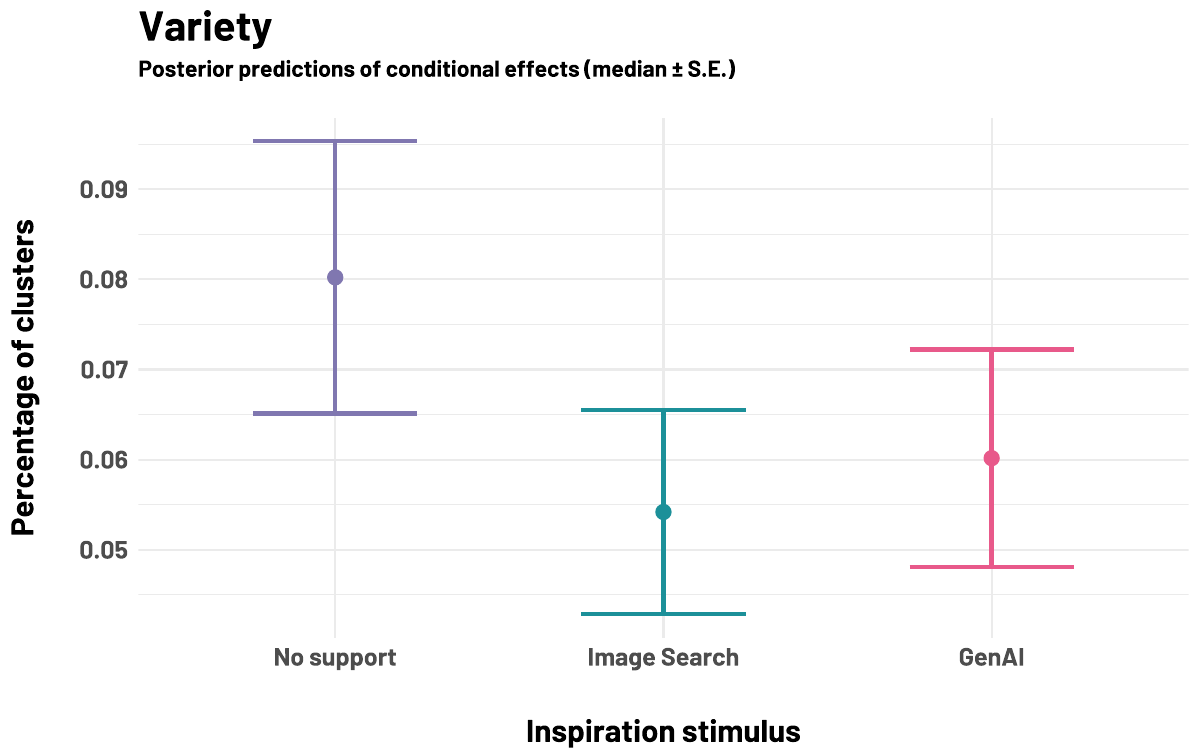}
    \caption{Model posterior predictions for \textsc{Variety} (percentage of clusters in which the participant has sketches). Error bars represent the standard error of the estimates.}
    \label{fig:variety_plot}
    \vspace{-1em}
\end{figure}

The total effects model shows a detrimental effect of both stimuli on \textsc{Variety}. The model suggests that the effect of \textsc{GenAI} has only an 86\% probability of being negative (mean = -.29, 89\% CI [-.75, .16]), and a Bayes Factor of 5.93 provides substantial evidence for the hypothesis that it yields a negative total effect on the variety of output. The effect of \textsc{Image Search} was even more negative (mean = -.40, [-.87, -.07]), with a Bayes Factor of 11.38 providing strong support for the hypothesis that it has a negative effect. 

The model including \textsc{Fluency} as a covariate, models the direct effect of the stimulus on the \textsc{Variety} of the output. Comparing the two models, we see that after accounting for the number of sketches that the participant produced, Google Image Search did not have much of an effect on \textsc{Variety} (mean = .02 [-.39, .43]. However, GenAI still had a negative effect (mean = -.15 [-55, .24]), with a Bayes Factor of 2.64 suggesting anecdotal evidence for this effect being negative.
\hl{In summary, \textbf{neither  \textsc{Image Search} nor \textsc{GenAI} provided meaningful support over the baseline in terms of enhancing the variety of the output, yielding, on average, lower variety than the baseline}}\textbf{. The effect of \textsc{Image Search} was fully mediated by \textsc{Fluency}, but \textsc{GenAI} also had an additional negative direct effect on \textsc{Variety}}.

\subsection{Originality}
To model \textsc{Originality}, we consider the number of other participants with sketches in the same cluster as each sketch.  
As in the case of \textsc{Variety}, our causal model considers two effects of the stimulus on \textsc{Originality}: a direct effect and an effect mediated by \textsc{Variety}.
We model these effects through two models, with and without \textsc{Variety} as a covariate.

\begin{table}[h]
\caption{Summary of the linear regression model for \textsc{Originality}---Direct Effects: \texttt{Originality $\sim$  Stimulus + Variety + (1|Participant ID)} and Total Effects:  \texttt{Originality $\sim$  Stimulus + (1|Participant ID)}. We provide the posterior means of parameter estimates (Est.), posterior standard deviations of these estimates (SD), and the bounds of their 89\% compatibility interval. We note that this is not the same as the frequentist confidence interval but a percentile of the posterior distribution. All parameter estimates converged with an ESS well above 1000 and an R-hat of 1.00.}
\small{
\begin{tabular}{lllll}
\toprule
& \multicolumn{2}{c}{\textbf{Direct Effects}} & \multicolumn{2}{c}{\textbf{Total Effects}}\\
\textbf{Parameter} & \textbf{Est. (SD)} & \textbf{89\% CI}  & \textbf{Est. (SD)} & \textbf{89\% CI} \\
\midrule
Intercept                           & .83 (0.02) & [.81, .86]& .86 (.01) & [.84, .88]\\
Image Search                        & -.01 (0.02) & [-.03, .02] & -.01 (.02) & [-.04, .01]\\
GenAI                               & -.03 (0.02) & [-.05, -.01] & -.03 (.02)& [-.06, -.01]\\
Variety                             & .01 (<0.01) & [.00, .01] &  & \\
\bottomrule
\end{tabular}
}
\label{tab:originality_model}
\vspace{-1em}
\end{table}

\begin{figure}[h]
\centering
    \includegraphics[width=0.425\textwidth]{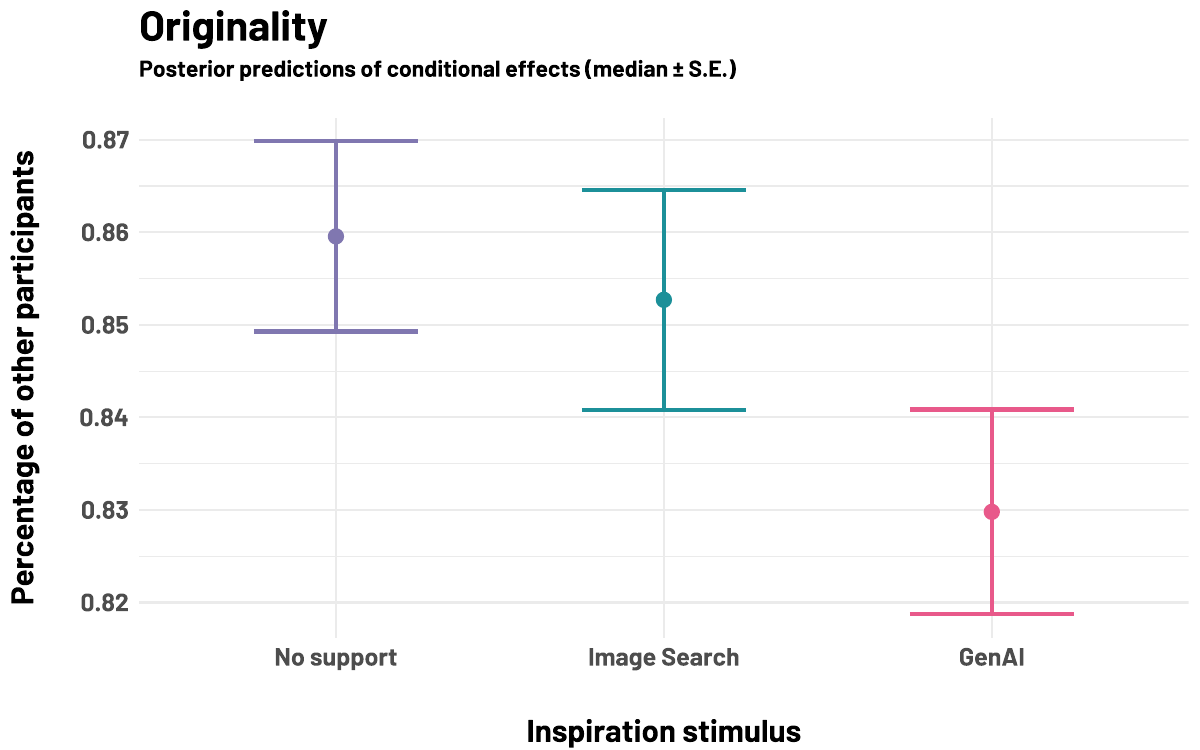}
    \caption{Model posterior predictions for originality (percentage of other participants who did not have an idea in the same cluster of an idea, averaged per participant). Error bars represent the standard error of the estimates.}
    \label{fig:originality_plot}
\end{figure}

The expected value for each response is based on a linear regression model. This models the originality score based on the inspiration stimulus (and variety score in the direct effects model), as well as a random effect of the participant. We use weakly informative, regularising priors for the model parameters (drawn from a normal distribution with mean zero and standard deviation of 2 for coefficients). We modelled our random effects as being drawn from a normal distribution with mean zero and standard deviation computed from the data through partial pooling. 

The total effects model suggests that both stimuli had small negative effects on \textsc{Originality}.
The model suggests that the effect of \textsc{GenAI} has a 97\% probability of being negative (mean = -.03, 89\% CI [-.06, .00]), and a Bayes Factor of 35. provides extreme evidence for the hypothesis that it yields a negative effect on the variety of output. The effect of \textsc{Image Search} was slightly less negative but also rather small (mean = -0.01, [-.04, .01]), but a Bayes Factor of 3.9 provides only anecdotal evidence against the hypothesis that it has a positive effect. Adding \textsc{Variety} did not change the model in any meaningful way, suggesting that \textsc{Variety} does not mediate an effect on \textsc{Originality}.
\hl{In summary, \textbf{neither \textsc{Image Search} nor \textsc{GenAI} provided a considerable aid in terms of developing \textsc{Originality} of the output, offering, on average, lower originality than the baseline, but these effects were negligible}.}

\begin{table*}[t]
    \centering
    \caption{\hl{Frequency of words used in the prompt by the participants in GenAI condition.}}
    \begin{tabular}{llllll}
    \toprule
         \textbf{Word}&  \textbf{Length}&  \textbf{Count}&  \textbf{Weighted percentage}&  \textbf{Similar words}& \textbf{Included in the brief}
\\
\midrule
         robot&  5&  38&  10.80\%&  robot, robots& No
\\
         kind&  4&  27&  7.67\% &  kind& Yes
\\
         chatbot&  7&  24&  6.82\%&  chatbot& Yes
\\
         intelligent&  11&  20&  5.68\%&  intelligent& Yes
\\
         cute&  4&  20&  5.68\%&  cute& No
\\
         caring&  6&  19&  5.40\%&  caring& Yes
\\
         loving&  6&  17&  4.83\%&  love, loving& Yes
\\
\bottomrule
    \end{tabular}
    \label{tab:prompt_frequency}
\end{table*}

\vspace{1em}

\subsection{Why did ideating with Generative AI cause design fixation?}

The results from our statistical models suggest that support from Generative AI led to higher design fixation. To understand why this occurred, this section draws on our interview data, the prompts created by participants, the AI-generated images, and the participants' sketches. We first explore the content of participants' prompts as one potential cause of design fixation. This encapsulates how participants claimed to develop the prompts and how they were influenced by the design brief or the example design. Next, we quantitatively explore the similarity between participants' sketches and the AI images used to inform that sketch in terms of design fixation. We then discuss the types of AI-generated images returned by participants' prompts and the sketches created based on them, using a case-study-based approach to illustrate our claims.  

\textbf{Overall, our analysis indicates that participants frequently relied on prompts containing keywords copied directly from the design brief or used prompts inspired by the example design. These prompts resulted in AI-generated images that were conceptually similar to the example design in 44\% of cases and which frequently contained fixating features that were present in the example design. Further, while not all sketches exhibit high similarity to the example we provided, ideating based on AI images can lead to \textit{fixation displacement}, where participants simply fixate on the images generated by the AI and copy what they see. This can occur irrespective of whether the participant imitates the example design or whether they attempt to explore other areas of the conceptual space}.

\subsubsection{``I just took the words from the brief'': Fixation from prompts based on the brief and example design}

One plausible source of design fixation in our experiment is the prompts that participants used to generate AI images. That is, if prompts include terms that are closely related to the example design or which draw from the design brief, then they might conceivably give rise to AI-generated images that are similar.

To investigate this possibility, we first analysed the prompts that participants used for generating images. 
Participants created a total of 117 prompts, with a mean of 5.85 prompts per participant (median = 5.5 range = 2--15). The length of each prompt ranged from 1 to 26 words (mean = 3.5 words). To explore the content of the prompts, we conducted a simple word frequency analysis using the automated word counting feature in NVivo12. This feature enables us to identify the total number and frequency of unique words that appear in the prompts.  Table~\ref{tab:prompt_frequency} shows a summary of the most frequent words appearing in participants' prompts.

This analysis revealed that participants frequently created prompts by using keywords copied from the design brief. In total, 52 prompts (44\%) contained at least one word that appears in the brief. Examples included \textit{kind}, which appeared in 27 different prompts; \textit{chatbot}, which appeared in 24 prompts;
 \textit{intelligent}, which appeared in 20 prompts; and \textit{caring}, which appeared in 19 prompts. 
During the interviews, participants reported using this approach due to a feeling of being `stuck' when trying to develop a prompt. Others attempted to generate ideas that met the requirements of the design brief. GenAI-P253, for example, described adapting content from the brief into his prompt and told us that the process he followed was to ``\textit{
read the brief, take the descriptions that they had, and make sure that I was meeting those descriptions.}''

A second approach involved participants using keywords that were themed around the example design. In total, 57\% (67/117) of the prompts contained the word \textit{robot}, \textit{chatbot} or \textit{chatbox} (a homophone of chatbots).

This suggests that participants often translated what they saw into a prompt before ideating based on the results. In addition, 78 prompts (66.6\%) included terms related to robots alongside words from the design brief. For example, P437's very first prompt was \textit{`kind loving caring robot'} whereas GenAI-P253 entered \textit{`cute kind chatbox character'}.

Data from the interviews also supports the notion that participants created prompts which were fixated on the example design. GenAI-P605, for example, described their process as starting with `robots' and then trying to factor in other aspects of the design brief. He said that he, ``\textit{searched up intelligent robots. But all those robots that I saw in the [AI-generated images], they looked intelligent, but they didn't look kind or caring. [I thought], how can I make it both caring and intelligent?}''. This participant created three distinct prompts: \textit{Intelligent robot}, \textit{Caring robot}, and \textit{Baymax} (referring to an inflatable computerised robot from a Disney movie) in an attempt to come up with alternative ideas.  

However, it is worth noting participants created 39 prompts that did not contain words from the brief or phrases related to robots. These prompts evince participants' attempts to explore different possibilities within the conceptual space of a `kind and loving' character. GenAI-P166, for example, recounted how they started the task by reading the brief and thinking about what to draw. This led them to the idea of `family', which they then translated into three distinct but related prompts: \textit{family}, \textit{mom}, and \textit{Mom - young}. They subsequently drew a sketch of a woman's face as their only design after seeing the images Midjourney returned from these prompts. 

Taken together, these cases illustrate how creating prompts based on the brief and the example design may be an initial stimulus for fixation. A successful strategy to overcome this problem was to try to think `beyond' the brief and the example. This latter quality is what may be needed from AI systems that truly support designers in avoiding fixation.

\subsubsection{``I would just kind of copy it and then tweak'': AI-generated images as a cause of fixation}

A second putative cause of design fixation in our experiment is the AI imagery that participants saw. That is, if the AI images were not meaningfully different to the example design, then this may have encouraged fixation because participants did not consider (or simply were not exposed to) other possible alternatives. This explanation is plausible given that 66.6\% of all prompts contained terms related to robots or words copied from the design brief. Prompts containing these terms might be expected to produce images similar to the example design, in turn leading to fixated sketches.

To explore the relationship between fixation in participants' sketches and the AI images, we first computed the correlation between the design fixation score of participants' sketches (previously calculated by two independent raters, see Section \ref{sec:materials}) and the design fixation score of the most recent set of AI-generated images immediately preceding each sketch. We selected Spearman’s rank correlation (a non-parametric test) as the data did not satisfy normality assumptions. 

For this analysis, we begin with the total set of sketches produced by participants in the AI condition (92 in total). We found that 10 of these sketches were created prior to entering any prompts into Midjourney; we, therefore, removed these sketches from consideration as there are no equivalent AI images to compare them against. This left us with 82 sketches, which we plotted against the relevant AI-generated images seen immediately before drawing the sketch. 

\begin{figure}[H]
\centering
    \includegraphics[width=0.5\textwidth]{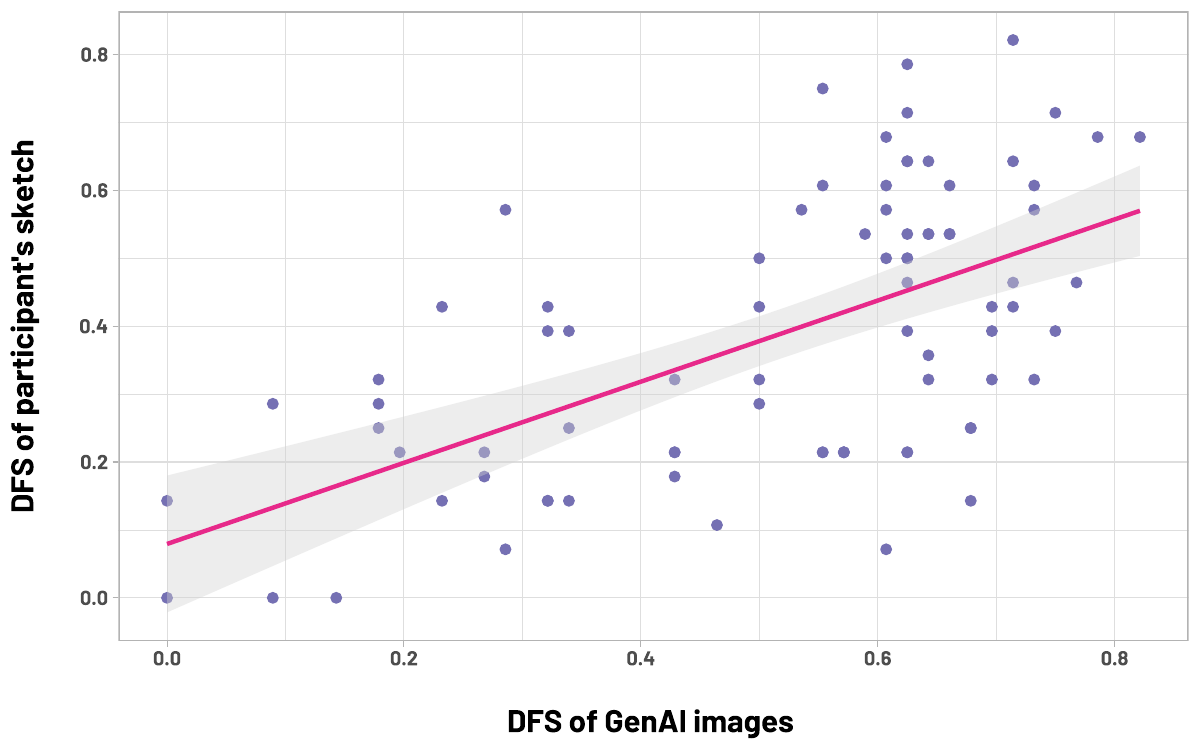}
    \caption{Scatterplot illustrating the correlation between the design fixation score of GenAI images appearing immediately before a participant's sketch and the design fixation score of the sketch associated with the same set of GenAI images. DFS = design fixation score.}
    \label{fig: copiedfromAI}
\end{figure}

Figure \ref{fig: copiedfromAI} illustrates the correlation. We observed a moderate positive correlation between the design fixation score of each sketch and the design fixation score of the AI images immediately preceding that sketch ($\rho$ = 0.56). This provides quantitative support for the idea that \textbf{AI-generated images that contained features of the example avatar led to sketches with higher design fixation scores}.

\begin{figure*}[h!]
\centering
\includegraphics[width=0.575\textwidth]{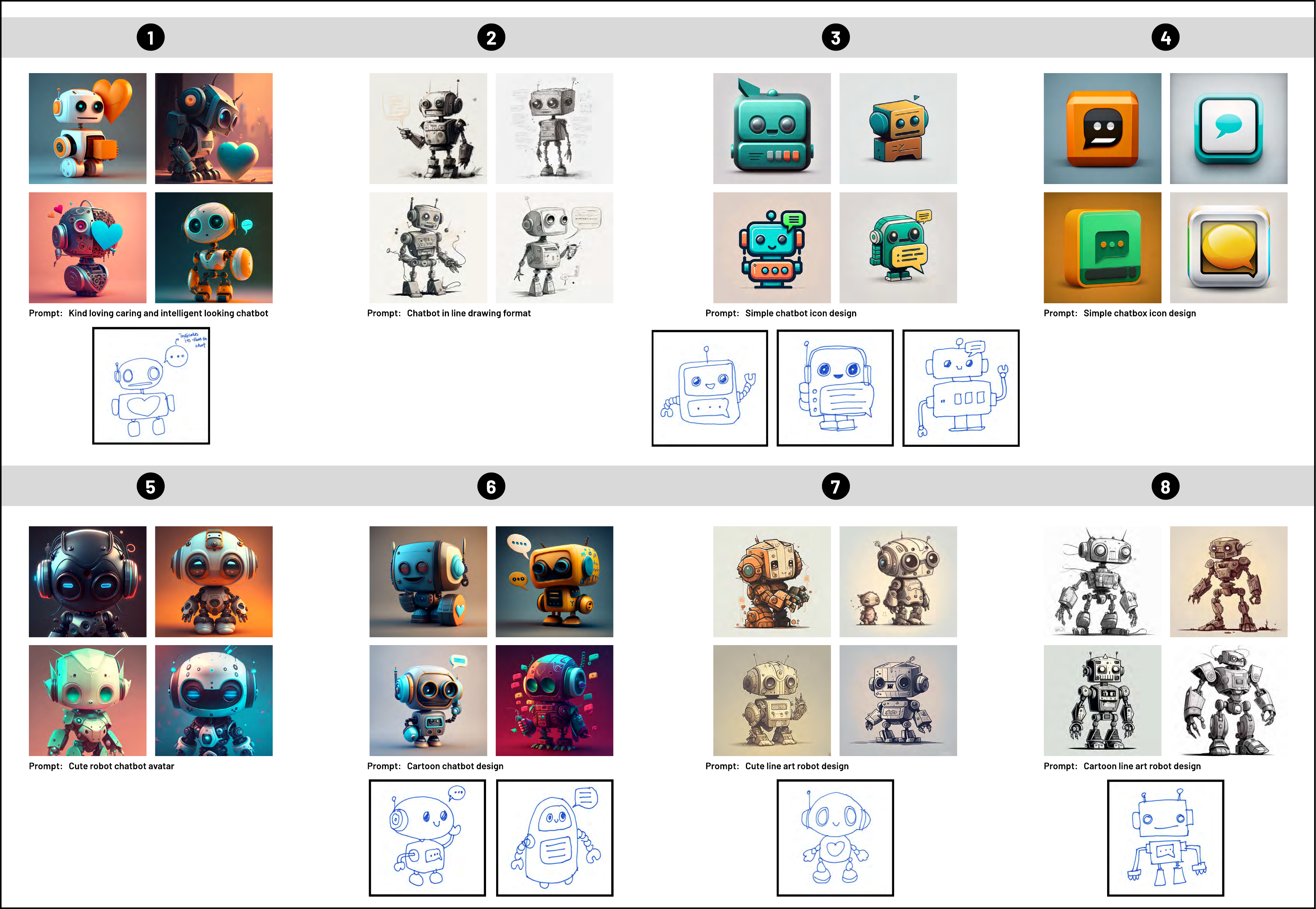}
\caption{An example of a participant producing sketches based on AI images that are similar to the example design, evidencing fixation when co-ideating with AI.}
\label{fig: cascading_approach_sample}
\end{figure*}

Next, we qualitatively investigated what kinds of images were generated by Midjourney in response to participants' prompts and whether the resulting sketches were fixated on these images. We began with a simple visual inspection of the AI images to probe whether Midjourney's outputs were meaningfully different to the example design. 

This inspection revealed that 44\% (206/468) of the AI-generated images portrayed humanoid robots that were conceptually similar to the example avatar. In turn, these images were qualitatively similar to the sketches participants made in response to them, indicating a tendency among participants to imitate --- or even directly copy --- what they saw.

Figure \ref{fig: cascading_approach_sample} shows an example of this phenomenon. The figure shows the AI images seen by one participant in Midjourney over time. It then positions the participant's sketches according to the most recently issued group of AI images before the sketch was drawn. In this example, it is immediately evident that the majority of sketches are superficially similar to the images returned by Midjourney. Likewise, these sketches are similar to the example design we provided (i.e. a cutesy robot) and typically contain the same salient features (legs, arms, and so on). The presence of these features and their inclusion in the subsequent sketches is one plausible explanation for why the AI support did not encourage participants to `break free' of fixation. It appears to have merely reinforced the existing problem.

\begin{figure*}[htp]
\centering
\includegraphics[width=0.575\textwidth]{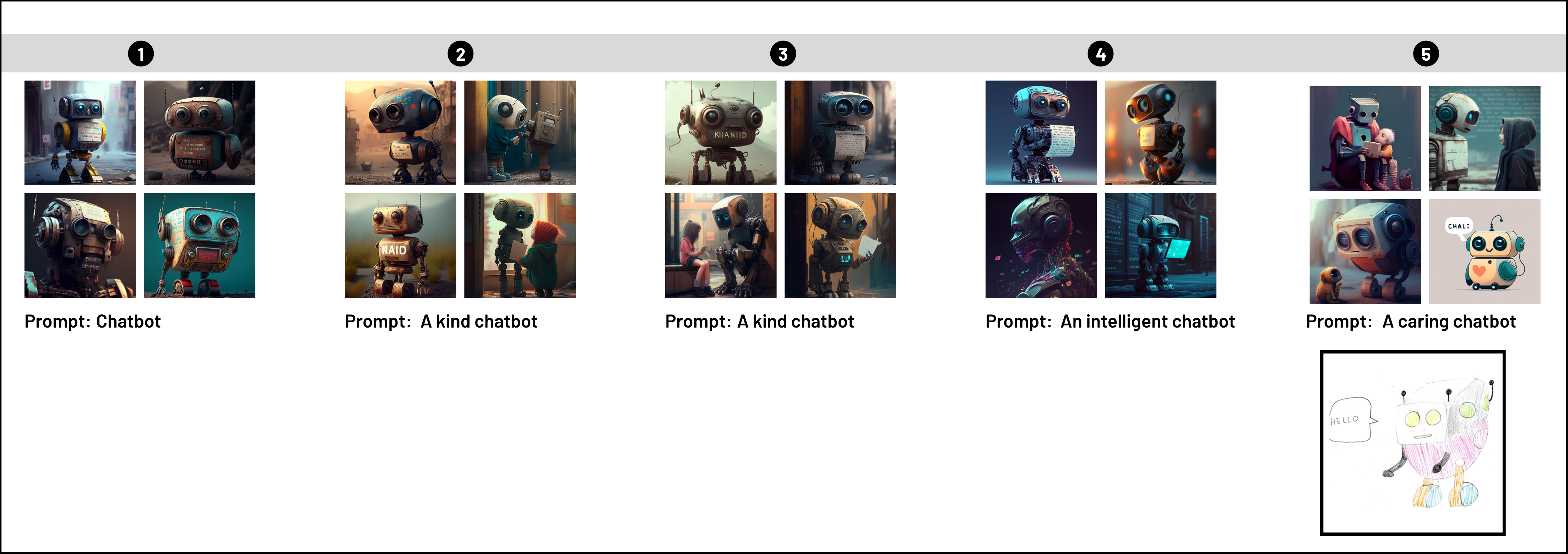}
\caption{A second example of a fixated sketch created after several rounds of prompting and image generation. In this case, the prompt is also fixated on the idea of a chatbot, creating similar returns from Midjourney each time.}
\label{fig: looped generationsmple}
\end{figure*}

This phenomenon arose irrespective of whether participants ideated on the fly or considered multiple ideas before creating a sketch. Figure \ref{fig: looped generationsmple} illustrates a second case where the participant is once again fixated on the idea of a robot. In this instance, the participant delays sketching until after issuing multiple prompts and seeing several rounds of AI images. It can be seen that the single sketch the participant created is of a robot-type character, evidencing fixation. The sequence also highlights how the prompt plays a role in this effect, with the participant attempting to vary their initial `chatbot' prompt by adding keywords such as `intelligent' or `kind', but resulting in thematically similar returns each time.

Cases such as these are illustrative of how fixation may have occurred, with participants repeatedly generating ideas that were similar to the example design and imitating the ideas within them. The interview data supported this latter idea. When asked about their approach to ideation, participants described the AI as a ``source of inspiration'' but openly admitted they sometimes copied what they observed. For example, GenAI-P253 claimed that using AI ``\textit{helped a lot of the inspiration for a lot of the designs that very much I just put down what I wanted it to give me, and I would just kind of copy it and then tweak it a little bit for the designs.} ''

Overall, these cases highlight the risk of AI simply reinforcing the phenomenon of fixation on an initial example. In turn, they raise the question of how AI systems might be usefully designed to encourage shifts away from this effect.

\subsubsection{The notion of ``Fixation Displacement''}

In addition to investigating how fixated sketches resulted from fixated images, our inspection of the sketches in relation to AI images revealed an additional phenomenon not well-captured by the correlational analysis. That is, there is evidence of what we describe as \textit{fixation displacement}, where the participant creates sketches with little relation to the original design but which are very clearly fixated on the AI imagery. Here, the sketches produced are both objectively and subjectively different to the example design but demonstrate a high degree of fixation with the AI images.

\begin{figure*}[htp]
\centering
\includegraphics[width=0.575\textwidth]{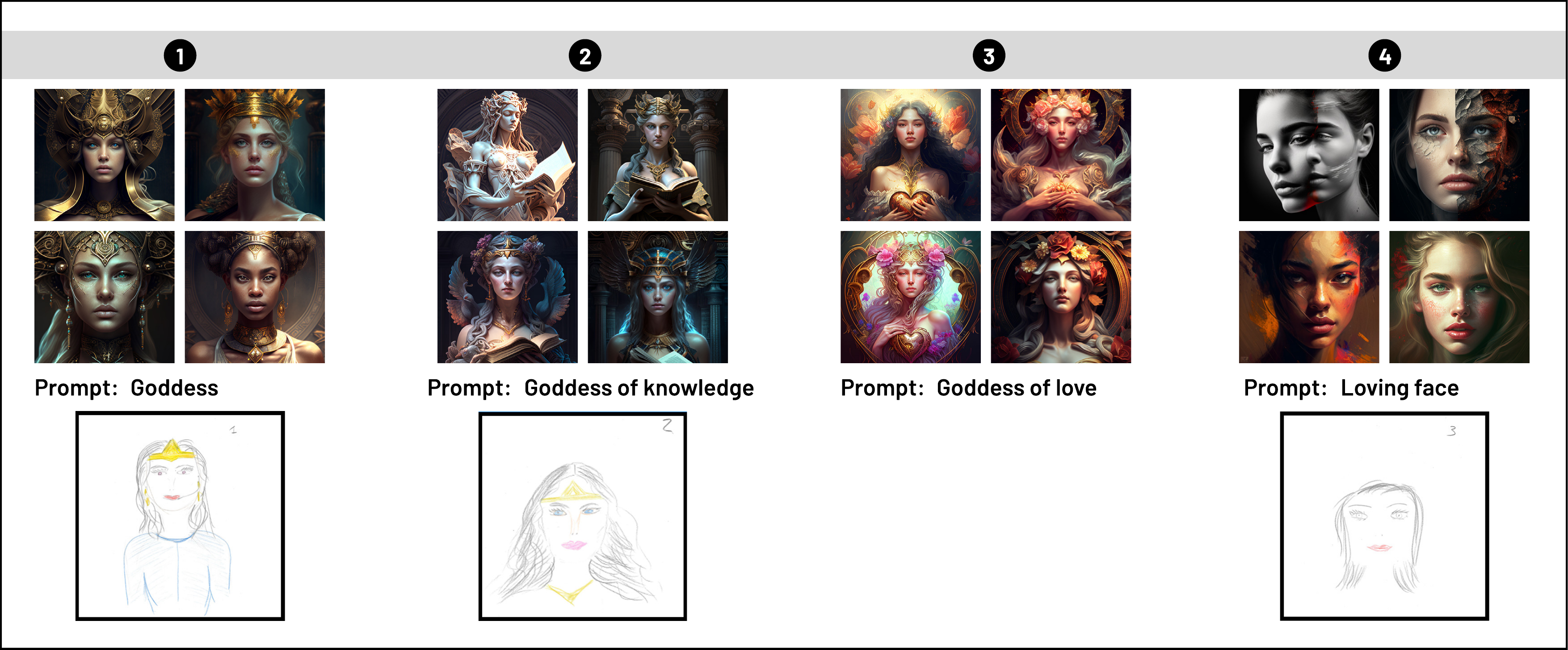}
\caption{An example of fixation displacement, this is where the participant has shifted their sketches away from the example robot avatar but has now become fixated on the idea of a woman's face via the AI images.}
\label{fig:fixation_displacement}
\vspace{-0.5em}
\end{figure*}

Figure \ref{fig:fixation_displacement} illustrates an example of fixation displacement in action. Here, the participant entered the prompt `goddess' as a way of beginning their ideation. This prompt has little connection to the design brief or the idea of a robot avatar. The participant then produced a sketch of a woman's face, which shares a small number of features with the example robot avatar (eyes, mouth, ears) but which is qualitatively different. Then, they proceed to iterate on this idea, resulting in three sketches that are similar in appearance and which bear a close resemblance to what the participant is seeing in Midjourney. 

\begin{figure*}[t]
\centering
\includegraphics[width=.575\textwidth]{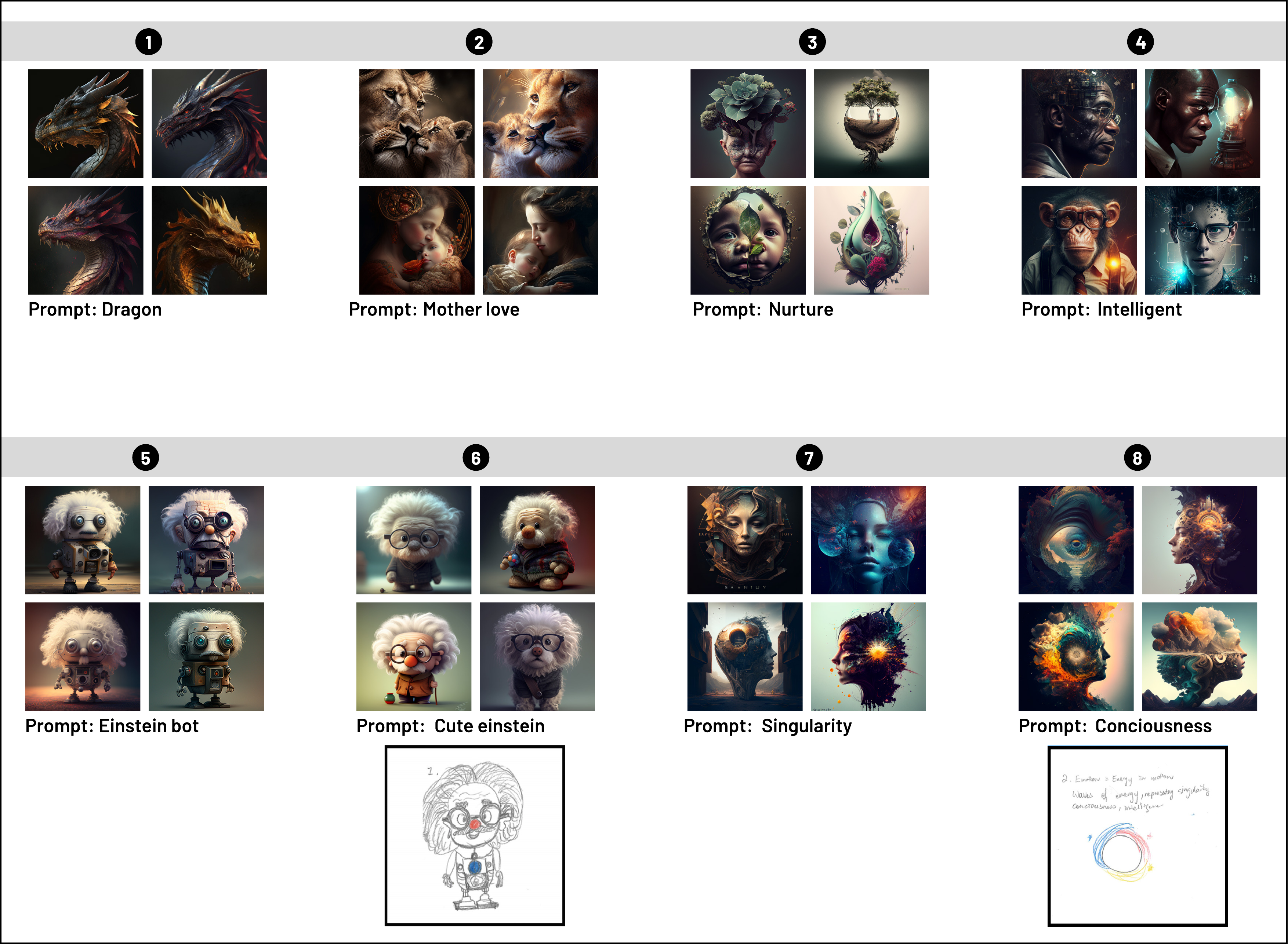}
\caption{An example of a participant progressing through different ideas and arriving at a final sketch that bears no resemblance to the example robot avatar.}
\label{fig:abstract_design}
\end{figure*}

Crucially, this phenomenon is not captured in our earlier scatterplot (Figure \ref{fig: copiedfromAI}) because the images and sketches have only a few features in common with the example design. This means they would be rated as quantitatively `low' on design fixation. In our experiment, fixation is operationalised in terms of similarity to the example design, where similarity is assessed by the presence or absence of features from the robot avatar. Conceptually, however, fixation refers to ``blind adherence to a set of ideas or concepts'' \cite{Jansson1991DesignFixation}. This general phenomenon is clearly depicted by the images and sketches in Figure~\ref{fig:fixation_displacement}, highlighting a new and novel risk of employing AI in ideation. That is, design fixation towards an initial example may not be overcome by using AI but may simply be displaced onto the examples that the AI provides. If one operationalises fixation in terms of deviation from an initial example, one might argue that such an outcome is apposite or even desired. But if one operationalises fixation in terms of blind adherence to an idea, then this outcome is questionable. 

What one may wish to see from AI-based ideation might be more akin to the process seen in Figure \ref{fig:abstract_design}. Here, the participant-generated 8 groups of images from Midjourney, beginning with prompts (such as `dragon') that have no relationship with the example design but which might inspire useful ideas and further exploration of the conceptual space. By the fourth prompt, the participant latched onto the idea of intelligence, which is then used to produce an Einstein-themed robot after prompt 6. However, the participant abandoned this idea and moved to an abstract design which bears no resemblance to the exemplar. While this still evidences some degree of fixation displacement, given the sketches the participant produced, it represents a significant conceptual shift from the example design. That is, the participant has considered a range of alternatives and has produced a seemingly useful design that bears no resemblance to the given example. This is perhaps more indicative of what we would consider to be effective AI-supported ideation.

\section{Discussion}

\begin{figure}[h]
    \centering
    \includegraphics[width=0.5\textwidth]{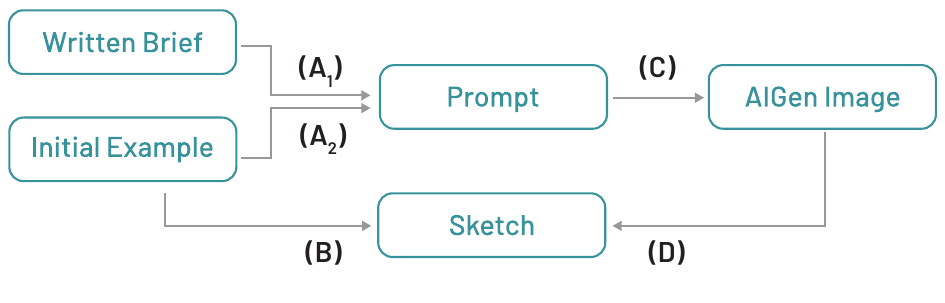}
    \caption{The overall ideation workflow of participants in the AI-supported condition,\textbf{($A_1$)}: Written brief to the prompt,\textbf{($A_2$)}: Initial example to the prompt,\textbf{($B$)}: Initial example to the sketch, \textbf{($C$)}: Prompt to AI-generated images, \textbf{($D$)}: AI-generated images to the sketch}
    \label{fig: discussion}
    \vspace{-1.05em}
\end{figure}

 This study aimed to identify the effects of using an AI image generator as inspiration support for an ideation task. Our quantitative analysis revealed that using AI-generated images had a detrimental effect on participants' ideation performance. Therefore, we aimed to uncover the cause of this effect through our qualitative analysis. We identified that AI caused more design fixation in participants and hindered the variety, originality and fluency of ideas compared to the baseline condition. Further, we observed a moderate positive correlation between the design fixation score of participants' sketches and the design fixation score of AI-generated images, which suggests that AI has a potential influence in determining the outcome of the ideas.  Further, we observed that AI induced a fixation displacement in participants where, even if they shifted their focus away from the initial example, they became fixated on the AI-generated images instead. In this section, we reflect on our learnings and discuss potential opportunities for developing generative AI to better facilitate ideation tasks, and propose strategies for improving divergent thinking during ideation. In doing so, we look at different phases of the ideation task performed by participants in detail (see figure \ref{fig: discussion}). 

\subsection{How to avoid fixation on the design brief when determining prompts for Generative AI?}

Through our study, we identified that the prompt acted as a potential source of design fixation. While participants used different strategies to devise their prompts (Figure\ref{fig: discussion}-A), the results suggest that most of them used keywords from the brief (Figure\ref{fig: discussion}-A$_1$) and built upon the idea of a `robot' from the initial example (Figure\ref{fig: discussion}-A$_2$) when devising their prompts. Participants claimed that they tried to avoid copying the initial example later during sketching (Figure\ref{fig: discussion}-B), but this was not reflected in the data given that the high design fixation score was calculated based on the ratio of replicated features from the initial example. Therefore, we assume that poor prompt design led to the generated images sharing similar features with the example. This suggests that design fixation happened when the participants created the prompts (Figure\ref{fig: discussion}-A). Participants tended to produce prompts that were semantically similar to the words given in the design brief and examples, while exhibiting strategies of repeating the same steps when creating prompts. Prior work indicates that participants can become fixated on exposed examples~\cite{Jansson1991DesignFixation, Cardoso2009DesignProblem-solving}. Further, studies have shown that participants tend to fixate on self-generated ideas and concepts compared to initial examples~\cite{Leahy2020DesignIdeas}. Our study aligns with these findings, suggesting that some participants were fixated on the design brief, the example avatar, or self-generated ideas when determining keywords for the prompt. 

Thus, paying attention to different prompting strategies might help mitigate this first potential occurrence of fixation when co-ideating with AI. Youmans et al.~\cite{Youmans2014DesignPrevention} summarise different strategies to mitigate design fixation based on the cause of occurrence. One strategy to snap participants out of fixation is to have timely warnings to consider alternative options~\cite{Youmans2014DesignPrevention}. Therefore, creativity support systems based on generative AI could not only turn prompts into images but also scaffold users' abilities to craft better prompts for ideation. AI systems could support users in creatively interpreting the brief, push users' thinking into alternative directions, or mix arbitrary ideas into the prompts. This functionality could be enabled through other generative AI techniques, such as large language models. 

Cheng et al.~\cite{Cheng2014ADesigners} have found that showing low-fidelity, abstract, and partially completed ideas led participants to become more divergent in their thinking and reduce fixation. Users of AI systems should consider using prompts that generate low-fidelity abstract or partial images when interacting with AI because images with these qualities might alleviate design fixation and encourage divergent thinking in an ideation task. Having a predetermined prompt structure or template that describes ways to make the images more abstract and less refined might lower the risk of fixation. 

\subsection{How can AI generate images that better support ideation?}

The images generated by the AI system in this study were high in fidelity, visual detail, and quality; they appeared to be rich in shape, form, texture, colour, composition, and visual expressiveness (see Figure \ref{fig: cascading_approach_sample}-\ref{fig:abstract_design}). Though this showcases impressive functionality, it might have amplified conformity towards the generated image, causing fixation displacement. This aligns with prior studies, which have shown that complete and strong examples carry the potential to cause fixation~\cite{Cardoso2009DesignProblem-solving, Smith1993ConstrainingIn, Chrysikou2005FollowingTask.,Cheng2014ADesigners}. 
Previous work has found that introducing some incubation time can help dissolve concrete exemplars into more abstract concepts, lowering fixation~\cite{Youmans2011DesignFixation, Smith2011AFixation} and supporting the emergence of novel ideas~\cite{Deo2021IdeaEffectiveness, Sio2009DoesReview}. Though this was not possible in our study, given the short time available for the task, it is a process that users of AI systems can incorporate into their ideation process. Further, when developing generative AI to support ideation, it may be useful to introduce mechanisms to lower the fidelity and the richness in detail of the output. Another direction is to show partially completed or blurred outputs, which might be beneficial for introducing ambiguity and pushing ideas in new directions~\cite{Cheng2014ADesigners}. Recent works by Davis et al.~\cite{Davis2019CreativeCreativity} and Williford et al.~\cite{Williford2023ExploringIdeation}  provide initial evidence suggesting that these mechanisms might be plausible approaches to be embedded in generative AI.

\subsection{How to translate AI images into design ideas?}

Through our visual comparisons, we observed similarities between the sketches produced by participants and the AI-generated images, suggesting that participants had imitated and, in some instances, directly copied elements from the images generated by Midjourney. Further, we identified that regardless of whether participants ideated on the fly or considered multiple ideas before creating a sketch, they gravitated towards features of the images which Midjourney generated, leading to fixation. Copying elements from an example is the easiest way to result in fixated outputs~\cite{Jansson1991DesignFixation, Cardoso2009DesignProblem-solving}, and our findings were consistent with this. 

To successfully act as sources of inspiration, generative AI tools must encourage more strategies that are more effective than copying. Previous work has shown that techniques like visual analogy---identifying abstract correspondences between the images being generated and the solution being sought---can improve the ideation effectiveness of designers of all levels, including novices~\cite{Casakin1999ExpertiseEducation}. However, Casakin and Goldschmidt highlight that even though novices already have an inherent understanding of how visual analogy works, they must be shown how to do it well and how it can support problem-solving in design activities ~\cite{Casakin1999ExpertiseEducation}. Scaffolding these skills is a promising role for AI-based creativity support tools.
 
We observed lower fluency in the GenAI and Image search conditions. Because the time given to complete the task was the same in all conditions, there was an inherent trade-off between spending time producing ideas vs. seeking inspiration. Interacting with both the AI image generator and the web image search led to less time spent sketching. These results tally with the findings of Vishwanathan and Linsey~\cite{Viswanathan2012DesignCost}, who found that though physical prototyping techniques that required more effort led to higher quality ideas in an engineering design task, they also increased design fixation and lowered fluency. They hypothesise that this is due to a ``sunk-cost effect''---the higher the effort spent in a given direction, the harder it is to move into a different one. Participants who spent more time refining prompts and interacting with the AI also had worse ideation performance. Users of generative AI systems should be careful and deliberate in their approach when seeking inspiration from external stimuli like AI image generators to mitigate the risk of design fixation. 
Crilly~\cite{Crilly2015FixationDesigners} suggests that empowering designers to recognise and reflect upon fixating episodes might be beneficial in developing a less fixating co-ideation workflow with AI. Further, Neroni and Crilly~\cite{Neroni2021HowRisk} state that uncovering participants' fixation tendencies, which they call "demonstrated vulnerability", is an effective approach that can further strengthen participants' ability to overcome fixation.  
In summary, when developing generative AI for co-ideation tasks, there is a rich opportunity for designing interactions with intelligent agents that not only generate stimuli but also encourage better ideation behaviours.
Triggering timely reminders, suggesting new idea directions, preempting fixation, varying the abstraction of the visual outcomes, and facilitating visual analogical reasoning are all promising directions for future work.

\subsection{Limitations}

We acknowledge several limitations in our study. First, at the time of writing, generative AI tools are still nascent technologies. Interaction paradigms are emerging, and users are still learning to leverage their potential. As such, our results describe a picture of somewhat na\"ive use of these tools. It will be interesting to see how these results evolve as users become accustomed to generative AI tools and incorporate them into their practice. 

Next, in this study, we gave all groups of participants the same amount of time to complete the task, but we observed lower fluency in the \textsc{GenAI} condition. We note that when the experiment took place, the AI system did not produce results instantaneously, which potentially delayed participants in that condition. However, we also note that participants in the \textsc{Image Search} condition, who \textit{did} get their results instantaneously, also exhibited lower fluency than the baseline. This could be due to the exposure to a large number of images with endless scrolling, which added another layer of decision making to pick the ideas that suit them. Therefore, in a real-world setting, it is important to consider the trade-off between spending time on the task (e.g. by sketching) vs. seeking inspiration (e.g. by interacting with a creativity-support tool).

We acknowledge that because we limited the task time to 20 minutes based on previous work, we restrict the scope of our insights to short-term usage of these tools in a rapid ideation task. In the real world, people may spend longer reflecting on the outputs of AI, and incubation time along with iteration of sketched ideas may produce results different from those of our experiment. Further, we screened participants for prior skills in visual design, but few had professional industry experience. Though our sample was balanced across conditions, we make no claims about how these effects might be affected by expertise.Therefore, with this study, we can only provide initial insights into how a novice designer might approach a design task, and to generalize these claims, we need further investigation.

In this study, we operationalised design fixation by looking for a restricted set of salient features from the example in participants' sketches. Though the choice of focusing on denotative elements of the design aims to facilitate operationalization, we acknowledge that there are also connotative aspects that were left outside the scope of our analysis, including art style, emotional expression, and cultural references. Finally, our study only evaluated the potential of generative AI tools for ideation support through the specific example of image generators in a visual ideation task. It remains to be seen how these effects translate to other modalities, such as text, video, audio, and music generation.

\section{Conclusion}

Through this study, we contribute empirical evidence to the discussion of the potential of generative AI to augment human creativity. Our study revealed that using an AI image generator as a source of inspiration by novice designers led to higher design fixation on an initial example and lower fluency, variety, and originality of ideas compared to using a conventional image search or no inspiration support. We suggest that fixation can happen in how the brief and the example influence the prompt given to the AI system, how the system translates it into images, and how the images inspire participants' ideas. All of these offer rich opportunities for re-design. Our work suggests that, at least in the current context of AI tool usage, given a fixed amount of time for a visual ideation task, this time is better spent sketching than seeking inspiration through AI. Our work suggests that generative AI tools aimed at supporting co-ideation should not only focus on generating stimuli but also on encouraging more effective ideation behaviours. We believe that incorporating well-thought-out methods and strategies into user practices and developing generative AI tools that can reduce common obstacles, such as design fixation and other creativity blockers, can maximise its potential to speed up the creative process and improve the quality of innovative design output.

\begin{acks}
This research is supported by the Rowden White Scholarship and the Melbourne Research Scholarship offered by the University of Melbourne. We also would like to thank Christian Davey at the Melbourne Statistical Consulting Platform for their support.
\end{acks}

\bibliographystyle{ACM-Reference-Format}
\bibliography{references}


\begin{thebibliography}{76}


\ifx \showCODEN    \undefined \def \showCODEN     #1{\unskip}     \fi
\ifx \showDOI      \undefined \def \showDOI       #1{#1}\fi
\ifx \showISBNx    \undefined \def \showISBNx     #1{\unskip}     \fi
\ifx \showISBNxiii \undefined \def \showISBNxiii  #1{\unskip}     \fi
\ifx \showISSN     \undefined \def \showISSN      #1{\unskip}     \fi
\ifx \showLCCN     \undefined \def \showLCCN      #1{\unskip}     \fi
\ifx \shownote     \undefined \def \shownote      #1{#1}          \fi
\ifx \showarticletitle \undefined \def \showarticletitle #1{#1}   \fi
\ifx \showURL      \undefined \def \showURL       {\relax}        \fi
\providecommand\bibfield[2]{#2}
\providecommand\bibinfo[2]{#2}
\providecommand\natexlab[1]{#1}
\providecommand\showeprint[2][]{arXiv:#2}

\bibitem[Alipour et~al\mbox{.}(2018)]%
        {Alipour2018AFactors}
\bibfield{author}{\bibinfo{person}{Leyla Alipour}, \bibinfo{person}{Mohsen Faizi}, \bibinfo{person}{Asghar Mohammad~Moradi}, {and} \bibinfo{person}{Gholamreza Akrami}.} \bibinfo{year}{2018}\natexlab{}.
\newblock \showarticletitle{{A review of design fixation: research directions and key factors}}.
\newblock \bibinfo{journal}{\emph{InternatIonal Journal of Design Creativity and Innovation}} \bibinfo{volume}{6}, \bibinfo{number}{2} (\bibinfo{year}{2018}), \bibinfo{pages}{22--35}.
\newblock
\showISSN{2165-0357}
\urldef\tempurl%
\url{https://doi.org/10.1080/21650349.2017.1320232}
\showDOI{\tempurl}


\bibitem[Andersson et~al\mbox{.}(2012)]%
        {Andersson2012DesignUnseen}
\bibfield{author}{\bibinfo{person}{Carina Andersson}, \bibinfo{person}{Yvonne Eriksson}, \bibinfo{person}{Lasse Frank}, {and} \bibinfo{person}{Bill Nicholl}.} \bibinfo{year}{2012}\natexlab{}.
\newblock \showarticletitle{{Design fixations among information design students: What has been seen cannot be unseen}}. In \bibinfo{booktitle}{\emph{DS 74: Proceedings of the 14th International Conference on Engineering {\&} Product Design Education (E{\&}PDE12) Design Education for Future Wellbeing,}}. \bibinfo{publisher}{Design Society}, \bibinfo{address}{Antwerp, Belguim}.
\newblock
\urldef\tempurl%
\url{https://www.designsociety.org/download-publication/33183/design_fixations_among_information_design_students_what_has_been_seen_cannot_be_unseen}
\showURL{%
\tempurl}


\bibitem[{Audi MediaCenter}(2022)]%
        {AudiMediaCenter2022ReinventingMediaCenter}
\bibfield{author}{\bibinfo{person}{{Audi MediaCenter}}.} \bibinfo{year}{2022}\natexlab{}.
\newblock \bibinfo{title}{{Reinventing the wheel? “FelGAN” inspires new rim designs with AI | Audi MediaCenter}}.
\newblock
\newblock
\urldef\tempurl%
\url{https://www.audi-mediacenter.com/en/press-releases/reinventing-the-wheel-felgan-inspires-new-rim-designs-with-ai-15097}
\showURL{%
\tempurl}


\bibitem[Bellows et~al\mbox{.}(2012)]%
        {Bellows2012TheFixation}
\bibfield{author}{\bibinfo{person}{B.G. Bellows}, \bibinfo{person}{J.F. Higgins}, \bibinfo{person}{M.A. Smith}, {and} \bibinfo{person}{R.J. Youmans}.} \bibinfo{year}{2012}\natexlab{}.
\newblock \showarticletitle{{The Effects of Individual Differences in Working Memory Capacity and Design Environment on Design Fixation}}.
\newblock \bibinfo{journal}{\emph{Proceedings of the Human Factors and Ergonomics Society Annual Meeting}}  \bibinfo{volume}{56} (\bibinfo{date}{9} \bibinfo{year}{2012}), \bibinfo{pages}{1977--1981}.
\newblock
\showISBNx{9780945289418}
\showISSN{10711813}
\urldef\tempurl%
\url{https://doi.org/10.1177/1071181312561293}
\showDOI{\tempurl}


\bibitem[Bellows et~al\mbox{.}(2013)]%
        {Bellows2013AnResearch}
\bibfield{author}{\bibinfo{person}{B.G. Bellows}, \bibinfo{person}{J.F. Higgins}, {and} \bibinfo{person}{R.J. Youmans}.} \bibinfo{year}{2013}\natexlab{}.
\newblock \showarticletitle{{An individual differences approach to design fixation: Comparing laboratory and field research}}. In \bibinfo{booktitle}{\emph{Design, User Experience, and Usability. Design Philosophy, Methods, and Tools. DUXU 2013. Lecture Notes in Computer Science}}. \bibinfo{publisher}{Springer}, \bibinfo{address}{Berlin Heidelberg}, \bibinfo{pages}{13--21}.
\newblock
\showISBNx{978-3-642-39229-0}
\urldef\tempurl%
\url{https://doi.org/10.1007/978-3-642-39229-0}
\showDOI{\tempurl}


\bibitem[Belski and Belski(2015)]%
        {Belski2015ApplicationExperts}
\bibfield{author}{\bibinfo{person}{Iouri Belski} {and} \bibinfo{person}{Ianina Belski}.} \bibinfo{year}{2015}\natexlab{}.
\newblock \showarticletitle{{Application of TRIZ in improving the creativity of engineering experts}}.
\newblock \bibinfo{journal}{\emph{Procedia Engineering}}  \bibinfo{volume}{131} (\bibinfo{year}{2015}), \bibinfo{pages}{792--797}.
\newblock
\urldef\tempurl%
\url{https://doi.org/10.1016/j.proeng.2015.12.379}
\showDOI{\tempurl}


\bibitem[Braun and Clarke(2022)]%
        {Braun2022ThematicGuide}
\bibfield{author}{\bibinfo{person}{Virginia Braun} {and} \bibinfo{person}{Victoria Clarke}.} \bibinfo{year}{2022}\natexlab{}.
\newblock \bibinfo{booktitle}{\emph{{Thematic analysis: a practical guide}}}.
\newblock \bibinfo{publisher}{SAGE Publications Inc.}, \bibinfo{address}{Thousand Oaks, California,United states}.
\newblock
\showISBNx{978-1-4739-5323-9}


\bibitem[Bürkner(2017)]%
        {burker2017brms}
\bibfield{author}{\bibinfo{person}{Paul-Christian Bürkner}.} \bibinfo{year}{2017}\natexlab{}.
\newblock \showarticletitle{{brms}: An {R} Package for {Bayesian} Multilevel Models Using {Stan}}.
\newblock \bibinfo{journal}{\emph{Journal of Statistical Software}} \bibinfo{volume}{80}, \bibinfo{number}{1} (\bibinfo{year}{2017}), \bibinfo{pages}{1--28}.
\newblock
\urldef\tempurl%
\url{https://doi.org/10.18637/jss.v080.i01}
\showDOI{\tempurl}


\bibitem[Cao et~al\mbox{.}(2021)]%
        {Cao2021UtilizingGeneration}
\bibfield{author}{\bibinfo{person}{J. Cao}, \bibinfo{person}{W. Zhao}, {and} \bibinfo{person}{X. Guo}.} \bibinfo{year}{2021}\natexlab{}.
\newblock \showarticletitle{{Utilizing EEG to Explore Design Fixation during Creative Idea Generation}}.
\newblock \bibinfo{journal}{\emph{Computational Intelligence and Neuroscience}}  \bibinfo{volume}{2021} (\bibinfo{year}{2021}).
\newblock
\urldef\tempurl%
\url{https://doi.org/10.1155/2021/6619598}
\showDOI{\tempurl}


\bibitem[Cardoso et~al\mbox{.}(2009)]%
        {Cardoso2009DesignProblem-solving}
\bibfield{author}{\bibinfo{person}{C. Cardoso}, \bibinfo{person}{P. Badke-Schaub}, {and} \bibinfo{person}{A. Luz}.} \bibinfo{year}{2009}\natexlab{}.
\newblock \showarticletitle{{Design fixation on non-verbal stimuli: The influence of simple vs rich pictorial information on design problem-solving}}. In \bibinfo{booktitle}{\emph{Proceedings of the ASME Design Engineering Technical Conference}}. \bibinfo{publisher}{ASME}, \bibinfo{address}{San Diego, California, USA.}, \bibinfo{pages}{995--1002}.
\newblock
\showISBNx{9780791849057}
\urldef\tempurl%
\url{https://doi.org/10.1115/DETC2009-86826}
\showDOI{\tempurl}


\bibitem[Carpenter et~al\mbox{.}(2017)]%
        {carpenter2017stan}
\bibfield{author}{\bibinfo{person}{Bob Carpenter}, \bibinfo{person}{Andrew Gelman}, \bibinfo{person}{Matthew~D Hoffman}, \bibinfo{person}{Daniel Lee}, \bibinfo{person}{Ben Goodrich}, \bibinfo{person}{Michael Betancourt}, \bibinfo{person}{Marcus~A Brubaker}, \bibinfo{person}{Jiqiang Guo}, \bibinfo{person}{Peter Li}, {and} \bibinfo{person}{Allen Riddell}.} \bibinfo{year}{2017}\natexlab{}.
\newblock \showarticletitle{Stan: A probabilistic programming language}.
\newblock \bibinfo{journal}{\emph{Journal of statistical software}}  \bibinfo{volume}{76} (\bibinfo{year}{2017}).
\newblock


\bibitem[Casakin and Goldschmidt(1999)]%
        {Casakin1999ExpertiseEducation}
\bibfield{author}{\bibinfo{person}{Hernan Casakin} {and} \bibinfo{person}{Gabriela Goldschmidt}.} \bibinfo{year}{1999}\natexlab{}.
\newblock \showarticletitle{{Expertise and the use of visual analogy: implications for design education}}.
\newblock \bibinfo{journal}{\emph{Design Studies}} \bibinfo{volume}{20}, \bibinfo{number}{2} (\bibinfo{date}{3} \bibinfo{year}{1999}), \bibinfo{pages}{153--175}.
\newblock
\showISSN{0142-694X}
\urldef\tempurl%
\url{https://doi.org/10.1016/S0142-694X(98)00032-5}
\showDOI{\tempurl}


\bibitem[Cheng et~al\mbox{.}(2014)]%
        {Cheng2014ADesigners}
\bibfield{author}{\bibinfo{person}{Peiyao Cheng}, \bibinfo{person}{Ruth Mugge}, {and} \bibinfo{person}{Jan~P.L. Schoormans}.} \bibinfo{year}{2014}\natexlab{}.
\newblock \showarticletitle{{A new strategy to reduce design fixation: Presenting partial photographs to designers}}.
\newblock \bibinfo{journal}{\emph{Design Studies}} \bibinfo{volume}{35}, \bibinfo{number}{4} (\bibinfo{year}{2014}), \bibinfo{pages}{374--391}.
\newblock
\showISSN{0142694X}
\urldef\tempurl%
\url{https://doi.org/10.1016/J.DESTUD.2014.02.004}
\showDOI{\tempurl}


\bibitem[Chiou et~al\mbox{.}(2023)]%
        {Chiou2023DesigningGenerators}
\bibfield{author}{\bibinfo{person}{Li-Yuan Chiou}, \bibinfo{person}{Peng-Kai Hung}, \bibinfo{person}{Rung-Huei Liang}, {and} \bibinfo{person}{Chun-Teng Wang}.} \bibinfo{year}{2023}\natexlab{}.
\newblock \showarticletitle{{Designing with AI: An Exploration of Co-Ideation with Image Generators}}. In \bibinfo{booktitle}{\emph{Proceedings of the 2023 ACM Designing Interactive Systems Conference}}. \bibinfo{publisher}{ACM}, \bibinfo{address}{New York, NY, USA}, \bibinfo{pages}{1941--1954}.
\newblock
\showISBNx{9781450398930}
\urldef\tempurl%
\url{https://doi.org/10.1145/3563657.3596001}
\showDOI{\tempurl}


\bibitem[Chrysikou and Weisberg(2005)]%
        {Chrysikou2005FollowingTask.}
\bibfield{author}{\bibinfo{person}{Evangelia~G Chrysikou} {and} \bibinfo{person}{Robert~W Weisberg}.} \bibinfo{year}{2005}\natexlab{}.
\newblock \showarticletitle{{Following the Wrong Footsteps: Fixation Effects of Pictorial Examples in a Design Problem-Solving Task.}}
\newblock \bibinfo{journal}{\emph{Journal of Experimental Psychology: Learning, Memory, and Cognition,}} \bibinfo{volume}{31}, \bibinfo{number}{5} (\bibinfo{year}{2005}), \bibinfo{pages}{1134--1148}.
\newblock


\bibitem[Chung(2022)]%
        {Chung2022ArtisticTools}
\bibfield{author}{\bibinfo{person}{John Joon~Young Chung}.} \bibinfo{year}{2022}\natexlab{}.
\newblock \showarticletitle{{Artistic User Expressions in AI-powered Creativity Support Tools}}. In \bibinfo{booktitle}{\emph{UIST 2022 Adjunct - Adjunct Proceedings of the 35th Annual ACM Symposium on User Interface Software and Technology}}. \bibinfo{publisher}{Association for Computing Machinery, Inc}, \bibinfo{address}{New York, NY, USA}, \bibinfo{pages}{1--4}.
\newblock
\showISBNx{9781450393218}
\urldef\tempurl%
\url{https://doi.org/10.1145/3526114.3558531}
\showDOI{\tempurl}


\bibitem[Crilly(2015)]%
        {Crilly2015FixationDesigners}
\bibfield{author}{\bibinfo{person}{Nathan Crilly}.} \bibinfo{year}{2015}\natexlab{}.
\newblock \showarticletitle{{Fixation and creativity in concept development: The attitudes and practices of expert designers}}.
\newblock \bibinfo{journal}{\emph{Design Studies}}  \bibinfo{volume}{38} (\bibinfo{date}{5} \bibinfo{year}{2015}), \bibinfo{pages}{54--91}.
\newblock
\showISSN{0142694X}
\urldef\tempurl%
\url{https://doi.org/10.1016/J.DESTUD.2015.01.002}
\showDOI{\tempurl}


\bibitem[Crilly and Cardoso(2017)]%
        {Crilly2017WhereDesign}
\bibfield{author}{\bibinfo{person}{Nathan Crilly} {and} \bibinfo{person}{Carlos Cardoso}.} \bibinfo{year}{2017}\natexlab{}.
\newblock \showarticletitle{{Where next for research on fixation, inspiration and creativity in design?}}
\newblock \bibinfo{journal}{\emph{Design Studies}}  \bibinfo{volume}{50} (\bibinfo{date}{5} \bibinfo{year}{2017}), \bibinfo{pages}{1--38}.
\newblock
\showISSN{0142694X}
\urldef\tempurl%
\url{https://doi.org/10.1016/J.DESTUD.2017.02.001}
\showDOI{\tempurl}


\bibitem[Davis et~al\mbox{.}(2019)]%
        {Davis2019CreativeCreativity}
\bibfield{author}{\bibinfo{person}{N. Davis}, \bibinfo{person}{S. Siddiqui}, \bibinfo{person}{P. Karimi}, \bibinfo{person}{M.L. Maher}, {and} \bibinfo{person}{K. Grace}.} \bibinfo{year}{2019}\natexlab{}.
\newblock \showarticletitle{{Creative sketching partner: A co-creative sketching tool to inspire design creativity}}. In \bibinfo{booktitle}{\emph{Proceedings of the 10th International Conference on Computational Creativity, ICCC 2019}}. \bibinfo{publisher}{Association for Computational Creativity}, \bibinfo{address}{North Carolina}, \bibinfo{pages}{358--359}.
\newblock
\showISBNx{9789895416011}


\bibitem[de~Bono(2008)]%
        {deBono2008SixHats}
\bibfield{author}{\bibinfo{person}{Edward de Bono}.} \bibinfo{year}{2008}\natexlab{}.
\newblock \bibinfo{booktitle}{\emph{{Six Thinking Hats}} (\bibinfo{edition}{revised edition} ed.)}.
\newblock \bibinfo{publisher}{Penguin}, \bibinfo{address}{United Kingdom}.
\newblock


\bibitem[Deo et~al\mbox{.}(2021)]%
        {Deo2021IdeaEffectiveness}
\bibfield{author}{\bibinfo{person}{Saurabh Deo}, \bibinfo{person}{Aimane Blej}, \bibinfo{person}{Senni Kirjavainen}, {and} \bibinfo{person}{Katja Holtta-Otto}.} \bibinfo{year}{2021}\natexlab{}.
\newblock \showarticletitle{{Idea Generation Mechanisms: Comparing the Influence of Classification, Combination, Building on Others, and Stimulation Mechanisms on Ideation Effectiveness}}.
\newblock \bibinfo{journal}{\emph{Journal of Mechanical Design, Transactions of the ASME}} \bibinfo{volume}{143}, \bibinfo{number}{12} (\bibinfo{date}{12} \bibinfo{year}{2021}), \bibinfo{pages}{1 -- 46}.
\newblock
\showISSN{10500472}
\urldef\tempurl%
\url{https://doi.org/10.1115/1.4051239/1109505}
\showDOI{\tempurl}


\bibitem[Eapen et~al\mbox{.}(2023)]%
        {Eapen2023HowCreativity}
\bibfield{author}{\bibinfo{person}{Tojin~T. Eapen}, \bibinfo{person}{Daniel~J. Finkenstadt}, \bibinfo{person}{Josh Folk}, {and} \bibinfo{person}{Lokesh Venkataswamy}.} \bibinfo{year}{2023}\natexlab{}.
\newblock \bibinfo{title}{{How Generative AI Can Augment Human Creativity}}.
\newblock
\newblock
\urldef\tempurl%
\url{https://hbr.org/2023/07/how-generative-ai-can-augment-human-creativity)}
\showURL{%
\tempurl}


\bibitem[Fiorineschi and Rotini(2023)]%
        {Fiorineschi2023UsesLiterature}
\bibfield{author}{\bibinfo{person}{Lorenzo Fiorineschi} {and} \bibinfo{person}{Federico Rotini}.} \bibinfo{year}{2023}\natexlab{}.
\newblock \showarticletitle{{Uses of the novelty metrics proposed by Shah et al.: what emerges from the literature?}}
\newblock \bibinfo{journal}{\emph{Design Science}}  \bibinfo{volume}{9} (\bibinfo{date}{5} \bibinfo{year}{2023}), \bibinfo{pages}{e11}.
\newblock
\showISSN{2053-4701}
\urldef\tempurl%
\url{https://doi.org/10.1017/DSJ.2023.9}
\showDOI{\tempurl}


\bibitem[Gero et~al\mbox{.}(1994)]%
        {Gero1994DesignAids}
\bibfield{author}{\bibinfo{person}{John Gero}, \bibinfo{person}{A~T Purcell}, \bibinfo{person}{J~S Gero}, \bibinfo{person}{H~M Edwards}, {and} \bibinfo{person}{E Matka}.} \bibinfo{year}{1994}\natexlab{}.
\newblock \showarticletitle{{Design fixation and intelligent design aids}}. In \bibinfo{booktitle}{\emph{Artificial Intelligence in Design ’94}}. \bibinfo{publisher}{Springer}, \bibinfo{address}{Dordrecht}, \bibinfo{pages}{483--495}.
\newblock
\urldef\tempurl%
\url{https://doi.org/10.1007/978-94-011-0928-4}
\showDOI{\tempurl}


\bibitem[Guilford(1956)]%
        {Guilford1956TheIntellect.}
\bibfield{author}{\bibinfo{person}{Joy~Paul Guilford}.} \bibinfo{year}{1956}\natexlab{}.
\newblock \showarticletitle{{The structure of intellect.}}
\newblock \bibinfo{journal}{\emph{Psychological bulletin}} \bibinfo{volume}{53}, \bibinfo{number}{4} (\bibinfo{year}{1956}), \bibinfo{pages}{267--293}.
\newblock
\showISSN{1939-1455}
\urldef\tempurl%
\url{https://doi.org/10.1037/h0040755}
\showDOI{\tempurl}


\bibitem[Hart and Staveland(1988)]%
        {Hart1988DevelopmentResearch}
\bibfield{author}{\bibinfo{person}{Sandra~G. Hart} {and} \bibinfo{person}{Lowell~E. Staveland}.} \bibinfo{year}{1988}\natexlab{}.
\newblock \showarticletitle{{Development of NASA-TLX (Task Load Index): Results of Empirical and Theoretical Research}}.
\newblock \bibinfo{journal}{\emph{Advances in Psychology}} \bibinfo{volume}{52}, \bibinfo{number}{C} (\bibinfo{date}{1} \bibinfo{year}{1988}), \bibinfo{pages}{139--183}.
\newblock
\showISSN{0166-4115}
\urldef\tempurl%
\url{https://doi.org/10.1016/S0166-4115(08)62386-9}
\showDOI{\tempurl}


\bibitem[Hoggenmueller et~al\mbox{.}(2023)]%
        {Hoggenmueller2023CreativeExplorations}
\bibfield{author}{\bibinfo{person}{Marius Hoggenmueller}, \bibinfo{person}{Maria~Luce Lupetti}, {and} \bibinfo{person}{Willem Van Der~Maden}.} \bibinfo{year}{2023}\natexlab{}.
\newblock \showarticletitle{{Creative AI for HRI Design Explorations}}. In \bibinfo{booktitle}{\emph{HRI '23: Companion of the 2023 ACM/IEEE International Conference on Human-Robot Interaction}}. \bibinfo{publisher}{Association for Computing Machinery}, \bibinfo{address}{New York, NY, USA}, \bibinfo{pages}{40--50}.
\newblock
\showISBNx{9781450399708}
\urldef\tempurl%
\url{https://doi.org/10.1145/3568294.3580035}
\showDOI{\tempurl}


\bibitem[Howard et~al\mbox{.}(2013)]%
        {Howard2013OvercomingMethods}
\bibfield{author}{\bibinfo{person}{Thomas Howard}, \bibinfo{person}{Anja Maier}, \bibinfo{person}{Balder Onarheim}, {and} \bibinfo{person}{Morten. Friis-Olivarius}.} \bibinfo{year}{2013}\natexlab{}.
\newblock \showarticletitle{{Overcoming design fixation through education and creativity methods}}. In \bibinfo{booktitle}{\emph{Proceedings of the International Conference on Engineering Design, ICED}}, Vol.~\bibinfo{volume}{7 DS75-07}. \bibinfo{publisher}{The Design Society}, \bibinfo{address}{Seoul Korea}, \bibinfo{pages}{139--148}.
\newblock
\showISBNx{9781904670506}
\urldef\tempurl%
\url{https://www.designsociety.org/download-publication/34578/overcoming_design_fixation_through_education_and_creativity_methods}
\showURL{%
\tempurl}


\bibitem[Hwang(2022)]%
        {Hwang2022TooProcesses}
\bibfield{author}{\bibinfo{person}{Angel Hsing~Chi Hwang}.} \bibinfo{year}{2022}\natexlab{}.
\newblock \showarticletitle{{Too Late to be Creative? AI-Empowered Tools in Creative Processes}}. In \bibinfo{booktitle}{\emph{Extended Abstracts of the 2022 CHI Conference on Human Factors in Computing Systems}}. \bibinfo{publisher}{Association for Computing Machinery}, \bibinfo{address}{New York, NY, USA}, \bibinfo{pages}{1--9}.
\newblock
\showISBNx{9781450391566}
\urldef\tempurl%
\url{https://doi.org/10.1145/3491101.3503549}
\showDOI{\tempurl}


\bibitem[Jansson and Smith(1991)]%
        {Jansson1991DesignFixation}
\bibfield{author}{\bibinfo{person}{David~G. Jansson} {and} \bibinfo{person}{Steven~M. Smith}.} \bibinfo{year}{1991}\natexlab{}.
\newblock \showarticletitle{{Design fixation}}.
\newblock \bibinfo{journal}{\emph{Design Studies}} \bibinfo{volume}{12}, \bibinfo{number}{1} (\bibinfo{date}{1} \bibinfo{year}{1991}), \bibinfo{pages}{3--11}.
\newblock
\showISSN{0142694X}
\urldef\tempurl%
\url{https://doi.org/10.1016/0142-694X(91)90003-F}
\showDOI{\tempurl}


\bibitem[Joon et~al\mbox{.}(2021)]%
        {Joon2021TheTools}
\bibfield{author}{\bibinfo{person}{John Joon}, \bibinfo{person}{Young Chung}, \bibinfo{person}{Shiqing He}, {and} \bibinfo{person}{Eytan Adar}.} \bibinfo{year}{2021}\natexlab{}.
\newblock \showarticletitle{{The Intersection of Users, Roles, Interactions, and Technologies in Creativity Support Tools; The Intersection of Users, Roles, Interactions, and Technologies in Creativity Support Tools}}. In \bibinfo{booktitle}{\emph{Designing Interactive Systems Conference 2021}}. \bibinfo{publisher}{ACM}, \bibinfo{address}{New York, NY, USA}, \bibinfo{pages}{1817 --1833}.
\newblock
\showISBNx{9781450384766}
\urldef\tempurl%
\url{https://doi.org/10.1145/3461778}
\showDOI{\tempurl}


\bibitem[Karimi et~al\mbox{.}(2020)]%
        {Karimi2020CreativeCo-creativityb}
\bibfield{author}{\bibinfo{person}{P. Karimi}, \bibinfo{person}{J. Rezwana}, \bibinfo{person}{S. Siddiqui}, \bibinfo{person}{M.L. Maher}, {and} \bibinfo{person}{N. Dehbozorgi}.} \bibinfo{year}{2020}\natexlab{}.
\newblock \showarticletitle{{Creative sketching partner: An analysis of human-AI co-creativity}}. In \bibinfo{booktitle}{\emph{International Conference on Intelligent User Interfaces, Proceedings IUI}}. \bibinfo{publisher}{Association for Computing Machinery}, \bibinfo{address}{New York, NY, USA}, \bibinfo{pages}{221--230}.
\newblock
\showISBNx{9781450371186}
\urldef\tempurl%
\url{https://doi.org/10.1145/3377325.3377522}
\showDOI{\tempurl}


\bibitem[Kay et~al\mbox{.}(2016)]%
        {kay2016bayes}
\bibfield{author}{\bibinfo{person}{Matthew Kay}, \bibinfo{person}{Gregory~L. Nelson}, {and} \bibinfo{person}{Eric~B. Hekler}.} \bibinfo{year}{2016}\natexlab{}.
\newblock \showarticletitle{Researcher-Centered Design of Statistics: Why Bayesian Statistics Better Fit the Culture and Incentives of HCI}. In \bibinfo{booktitle}{\emph{Proceedings of the 2016 CHI Conference on Human Factors in Computing Systems}} (San Jose, California, USA) \emph{(\bibinfo{series}{CHI '16})}. \bibinfo{publisher}{Association for Computing Machinery}, \bibinfo{address}{New York, NY, USA}, \bibinfo{pages}{4521–4532}.
\newblock
\showISBNx{9781450333627}
\urldef\tempurl%
\url{https://doi.org/10.1145/2858036.2858465}
\showDOI{\tempurl}


\bibitem[Kim et~al\mbox{.}(2013)]%
        {Kim2013ToIntentionality}
\bibfield{author}{\bibinfo{person}{Jieun Kim}, \bibinfo{person}{Hokyoung Ryu}, {and} \bibinfo{person}{Hyeonah Kim}.} \bibinfo{year}{2013}\natexlab{}.
\newblock \showarticletitle{{To Be Biased or Not to Be: Choosing between Design Fixation and Design Intentionality}}. In \bibinfo{booktitle}{\emph{CHI '13 Extended Abstracts on Human Factors in Computing Systems}} \emph{(\bibinfo{series}{CHI EA '13})}. \bibinfo{publisher}{Association for Computing Machinery}, \bibinfo{address}{New York, NY, USA}, \bibinfo{pages}{349--354}.
\newblock
\showISBNx{978-1-4503-1952-2}
\urldef\tempurl%
\url{https://doi.org/10.1145/2468356.2468418}
\showDOI{\tempurl}


\bibitem[Koch et~al\mbox{.}(2020)]%
        {Koch2020ImageSense:Partnerships}
\bibfield{author}{\bibinfo{person}{Janin Koch}, \bibinfo{person}{Nicolas Taffin}, \bibinfo{person}{Michel Beaudouin-Lafon}, \bibinfo{person}{Markku Laine}, \bibinfo{person}{Andrés Lucero}, {and} \bibinfo{person}{Wendy~E. MacKay}.} \bibinfo{year}{2020}\natexlab{}.
\newblock \showarticletitle{{ImageSense: An Intelligent Collaborative Ideation Tool to Support Diverse Human-Computer Partnerships}}.
\newblock \bibinfo{journal}{\emph{Proceedings of the ACM on Human-Computer Interaction}} \bibinfo{volume}{4}, \bibinfo{number}{CSCW1} (\bibinfo{date}{5} \bibinfo{year}{2020}), \bibinfo{pages}{27}.
\newblock
\showISSN{25730142}
\urldef\tempurl%
\url{https://doi.org/10.1145/3392850}
\showDOI{\tempurl}


\bibitem[Kozbelt and Durmysheva(2007)]%
        {Kozbelt2007UnderstandingPredictors}
\bibfield{author}{\bibinfo{person}{Aaron Kozbelt} {and} \bibinfo{person}{Yana Durmysheva}.} \bibinfo{year}{2007}\natexlab{}.
\newblock \showarticletitle{{Understanding Creativity Judgments of Invented Alien Creatures: The Roles of Invariants and Other Predictors*}}.
\newblock \bibinfo{journal}{\emph{The Journal of Creative Behavior}} \bibinfo{volume}{41}, \bibinfo{number}{4} (\bibinfo{date}{12} \bibinfo{year}{2007}), \bibinfo{pages}{223--248}.
\newblock
\showISSN{2162-6057}
\urldef\tempurl%
\url{https://doi.org/10.1002/J.2162-6057.2007.TB01072.X}
\showDOI{\tempurl}


\bibitem[Lamiroy and Potier(2022)]%
        {Lamiroy2022Lamuse:Inspiration}
\bibfield{author}{\bibinfo{person}{Bart Lamiroy} {and} \bibinfo{person}{Emmanuelle Potier}.} \bibinfo{year}{2022}\natexlab{}.
\newblock \showarticletitle{{Lamuse: Leveraging Artificial Intelligence for Sparking Inspiration}}.
\newblock \bibinfo{journal}{\emph{Lecture Notes in Computer Science (including subseries Lecture Notes in Artificial Intelligence and Lecture Notes in Bioinformatics)}}  \bibinfo{volume}{13221 LNCS} (\bibinfo{year}{2022}), \bibinfo{pages}{148--161}.
\newblock
\showISBNx{9783031037887}
\showISSN{16113349}
\urldef\tempurl%
\url{https://doi.org/10.1007/978-3-031-03789-4{\_}10/FIGURES/6}
\showDOI{\tempurl}


\bibitem[Leahy et~al\mbox{.}(2020)]%
        {Leahy2020DesignIdeas}
\bibfield{author}{\bibinfo{person}{Keelin Leahy}, \bibinfo{person}{Shanna~R. Daly}, \bibinfo{person}{Seda McKilligan}, {and} \bibinfo{person}{Colleen~M. Seifert}.} \bibinfo{year}{2020}\natexlab{}.
\newblock \showarticletitle{{Design fixation from initial examples: Provided versus self-Generated ideas}}.
\newblock \bibinfo{journal}{\emph{Journal of Mechanical Design, Transactions of the ASME}} \bibinfo{volume}{142}, \bibinfo{number}{10} (\bibinfo{date}{10} \bibinfo{year}{2020}), \bibinfo{pages}{101402}.
\newblock
\showISSN{10500472}
\urldef\tempurl%
\url{https://doi.org/10.1115/1.4046446/1074761}
\showDOI{\tempurl}


\bibitem[Lewis(2023)]%
        {Lewis2023AIxArtist:Block}
\bibfield{author}{\bibinfo{person}{Makayla Lewis}.} \bibinfo{year}{2023}\natexlab{}.
\newblock \bibinfo{title}{{AIxArtist: A First-Person Tale of Interacting with Artificial Intelligence to Escape Creative Block}}.
\newblock
\newblock


\bibitem[Linsey et~al\mbox{.}(2010)]%
        {Linsey2010AFaculty}
\bibfield{author}{\bibinfo{person}{J~S Linsey}, \bibinfo{person}{I Tseng}, \bibinfo{person}{K Fu}, \bibinfo{person}{J Cagan}, \bibinfo{person}{K~L Wood}, {and} \bibinfo{person}{C Schunn}.} \bibinfo{year}{2010}\natexlab{}.
\newblock \showarticletitle{{A Study of Design Fixation, Its Mitigation and Perception in Engineering Design Faculty}}.
\newblock \bibinfo{journal}{\emph{Journal of Mechanical Design (JMD)}} \bibinfo{volume}{132}, \bibinfo{number}{4} (\bibinfo{date}{4} \bibinfo{year}{2010}), \bibinfo{pages}{041003}.
\newblock
\urldef\tempurl%
\url{https://doi.org/10.1115/1.4001110}
\showDOI{\tempurl}


\bibitem[Lucero(2012)]%
        {Lucero2012FramingWork}
\bibfield{author}{\bibinfo{person}{Andrés Lucero}.} \bibinfo{year}{2012}\natexlab{}.
\newblock \showarticletitle{{Framing, Aligning, Paradoxing, Abstracting, and Directing: How Design Mood Boards Work}}. In \bibinfo{booktitle}{\emph{Proceedings of the Designing Interactive Systems Conference}}. \bibinfo{publisher}{Association for Computing Machinery}, \bibinfo{address}{New York, NY, USA}, \bibinfo{pages}{438--447}.
\newblock
\showISBNx{9781450312103}


\bibitem[Luchins(1942)]%
        {Luchins1942MechanizationEinstellung.}
\bibfield{author}{\bibinfo{person}{Abraham~S. Luchins}.} \bibinfo{year}{1942}\natexlab{}.
\newblock \showarticletitle{{Mechanization in problem solving: The effect of Einstellung.}}
\newblock \bibinfo{journal}{\emph{Psychological Monographs}} \bibinfo{volume}{54}, \bibinfo{number}{6} (\bibinfo{year}{1942}), \bibinfo{pages}{i--95}.
\newblock
\showISSN{0096-9753}
\urldef\tempurl%
\url{https://doi.org/10.1037/h0093502}
\showDOI{\tempurl}


\bibitem[Mazzone and Elgammal(2019)]%
        {Mazzone2019ArtIntelligence}
\bibfield{author}{\bibinfo{person}{Marian Mazzone} {and} \bibinfo{person}{Ahmed Elgammal}.} \bibinfo{year}{2019}\natexlab{}.
\newblock \showarticletitle{{Art, Creativity, and the Potential of Artificial Intelligence}}.
\newblock \bibinfo{journal}{\emph{Arts 2019, Vol. 8, Page 26}} \bibinfo{volume}{8}, \bibinfo{number}{1} (\bibinfo{date}{2} \bibinfo{year}{2019}), \bibinfo{pages}{26}.
\newblock
\showISSN{2076-0752}
\urldef\tempurl%
\url{https://doi.org/10.3390/ARTS8010026}
\showDOI{\tempurl}


\bibitem[McElreath(2020)]%
        {mcelreath2020statistical}
\bibfield{author}{\bibinfo{person}{Richard McElreath}.} \bibinfo{year}{2020}\natexlab{}.
\newblock \bibinfo{booktitle}{\emph{Statistical rethinking: A Bayesian course with examples in R and Stan (2e)}}.
\newblock \bibinfo{publisher}{Chapman and Hall/CRC}.
\newblock


\bibitem[Moreno et~al\mbox{.}(2016)]%
        {Moreno2016OvercomingFindings}
\bibfield{author}{\bibinfo{person}{Diana~P. Moreno}, \bibinfo{person}{Luciënne~T. Blessing}, \bibinfo{person}{Maria~C. Yang}, \bibinfo{person}{Alberto~A. Hern{\'{a}}ndez}, {and} \bibinfo{person}{Kristin~L. Wood}.} \bibinfo{year}{2016}\natexlab{}.
\newblock \showarticletitle{{Overcoming design fixation: Design by analogy studies and nonintuitive findings}}.
\newblock \bibinfo{journal}{\emph{AI EDAM}} \bibinfo{volume}{30}, \bibinfo{number}{2} (\bibinfo{date}{5} \bibinfo{year}{2016}), \bibinfo{pages}{185--199}.
\newblock
\showISSN{0890-0604}
\urldef\tempurl%
\url{https://doi.org/10.1017/S0890060416000068}
\showDOI{\tempurl}


\bibitem[Neroni and Crilly(2021)]%
        {Neroni2021HowRisk}
\bibfield{author}{\bibinfo{person}{Maria~Adriana Neroni} {and} \bibinfo{person}{Nathan Crilly}.} \bibinfo{year}{2021}\natexlab{}.
\newblock \showarticletitle{{How to Guard Against Fixation? Demonstrating Individual Vulnerability is More Effective Than Warning of General Risk}}.
\newblock \bibinfo{journal}{\emph{The Journal of Creative Behavior}} \bibinfo{volume}{55}, \bibinfo{number}{2} (\bibinfo{date}{6} \bibinfo{year}{2021}), \bibinfo{pages}{447--463}.
\newblock
\showISSN{2162-6057}
\urldef\tempurl%
\url{https://doi.org/10.1002/JOCB.465}
\showDOI{\tempurl}


\bibitem[Purcell and Gero(1996)]%
        {Purcell1996DesignFixation}
\bibfield{author}{\bibinfo{person}{A~Terry Purcell} {and} \bibinfo{person}{John~S Gero}.} \bibinfo{year}{1996}\natexlab{}.
\newblock \showarticletitle{{Design and other types of fixation}}.
\newblock \bibinfo{journal}{\emph{Design studies}}  \bibinfo{volume}{17} (\bibinfo{year}{1996}), \bibinfo{pages}{363--383}.
\newblock
\urldef\tempurl%
\url{https://doi.org/10.1016/S0142-694X(96)00023-3}
\showDOI{\tempurl}


\bibitem[Rafner et~al\mbox{.}(2023)]%
        {Rafner2023PictureProblem-Solving}
\bibfield{author}{\bibinfo{person}{Janet Rafner}, \bibinfo{person}{Blanka Zana}, \bibinfo{person}{Peter Dalsgaard}, \bibinfo{person}{Michael~Mose Biskjaer}, {and} \bibinfo{person}{Jacob Sherson}.} \bibinfo{year}{2023}\natexlab{}.
\newblock \showarticletitle{{Picture This: AI-Assisted Image Generation as a Resource for Problem Construction in Creative Problem-Solving}}. In \bibinfo{booktitle}{\emph{Proceedings of the 15th Conference on Creativity and Cognition}}. \bibinfo{publisher}{Association for Computing Machinery (ACM)}, \bibinfo{address}{New York, NY, USA}, \bibinfo{pages}{262--268}.
\newblock
\showISBNx{9781450383769}
\urldef\tempurl%
\url{https://doi.org/10.1145/3591196.3596823}
\showDOI{\tempurl}


\bibitem[Remy et~al\mbox{.}(2020)]%
        {Remy2020EvaluatingResearch}
\bibfield{author}{\bibinfo{person}{Christian Remy}, \bibinfo{person}{Lindsay Macdonald~Vermeulen}, \bibinfo{person}{Jonas Frich}, \bibinfo{person}{Michael~Mose Biskjaer}, {and} \bibinfo{person}{Peter Dalsgaard}.} \bibinfo{year}{2020}\natexlab{}.
\newblock \showarticletitle{{Evaluating creativity support tools in HCI research}}. In \bibinfo{booktitle}{\emph{DIS 2020 - Proceedings of the 2020 ACM Designing Interactive Systems Conference}}. \bibinfo{publisher}{Association for Computing Machinery, Inc}, \bibinfo{address}{New York, NY, USA}, \bibinfo{pages}{457--476}.
\newblock
\showISBNx{9781450369749}
\urldef\tempurl%
\url{https://doi.org/10.1145/3357236.3395474}
\showDOI{\tempurl}


\bibitem[Rosenkopf and Nerkar(2001)]%
        {Rosenkopf2001BeyondIndustry}
\bibfield{author}{\bibinfo{person}{Lori Rosenkopf} {and} \bibinfo{person}{Atul Nerkar}.} \bibinfo{year}{2001}\natexlab{}.
\newblock \showarticletitle{{Beyond local search: boundary-spanning, exploration, and impact in the optical disk industry}}.
\newblock \bibinfo{journal}{\emph{Strategic Management Journal}} \bibinfo{volume}{22}, \bibinfo{number}{4} (\bibinfo{date}{4} \bibinfo{year}{2001}), \bibinfo{pages}{287--306}.
\newblock
\showISSN{1097-0266}
\urldef\tempurl%
\url{https://doi.org/10.1002/SMJ.160}
\showDOI{\tempurl}


\bibitem[Sbai et~al\mbox{.}(2019)]%
        {Sbai2019DesIGN:Networks}
\bibfield{author}{\bibinfo{person}{Othman Sbai}, \bibinfo{person}{Mohamed Elhoseiny}, \bibinfo{person}{Antoine Bordes}, \bibinfo{person}{Yann LeCun}, {and} \bibinfo{person}{Camille Couprie}.} \bibinfo{year}{2019}\natexlab{}.
\newblock \showarticletitle{{DesIGN: Design inspiration from generative networks}}.
\newblock \bibinfo{journal}{\emph{Computer Vision – ECCV 2018 Workshops. ECCV 2018. Lecture Notes in Computer Science()}}  \bibinfo{volume}{11131} (\bibinfo{year}{2019}), \bibinfo{pages}{0--0}.
\newblock
\urldef\tempurl%
\url{https://doi.org/10.1007/978-3-030-11015-4}
\showDOI{\tempurl}


\bibitem[Schmettow(2021)]%
        {schmettow2021new}
\bibfield{author}{\bibinfo{person}{Martin Schmettow}.} \bibinfo{year}{2021}\natexlab{}.
\newblock \bibinfo{booktitle}{\emph{New statistics for design researchers}}.
\newblock \bibinfo{publisher}{Springer}.
\newblock


\bibitem[Shah et~al\mbox{.}(2003)]%
        {Shah2003MetricsEffectiveness}
\bibfield{author}{\bibinfo{person}{Jami~J. Shah}, \bibinfo{person}{Noe Vargas-Hernandez}, {and} \bibinfo{person}{Steve~M. Smith}.} \bibinfo{year}{2003}\natexlab{}.
\newblock \showarticletitle{{Metrics for measuring ideation effectiveness}}.
\newblock \bibinfo{journal}{\emph{Design Studies}} \bibinfo{volume}{24}, \bibinfo{number}{2} (\bibinfo{date}{3} \bibinfo{year}{2003}), \bibinfo{pages}{111--134}.
\newblock
\showISSN{0142-694X}
\urldef\tempurl%
\url{https://doi.org/10.1016/S0142-694X(02)00034-0}
\showDOI{\tempurl}


\bibitem[Shin et~al\mbox{.}(2023)]%
        {Shin2023IntegratingIdeation}
\bibfield{author}{\bibinfo{person}{Joon~Gi Shin}, \bibinfo{person}{Janin Koch}, \bibinfo{person}{Andrés Lucero}, \bibinfo{person}{Peter Dalsgaard}, {and} \bibinfo{person}{Wendy~E. MacKay}.} \bibinfo{year}{2023}\natexlab{}.
\newblock \showarticletitle{{Integrating AI in Human-Human Collaborative Ideation}}. In \bibinfo{booktitle}{\emph{Conference on Human Factors in Computing Systems - Proceedings}}. \bibinfo{publisher}{Association for Computing Machinery}, \bibinfo{address}{New York, NY, USA}, \bibinfo{pages}{1--5}.
\newblock
\showISBNx{9781450394222}
\urldef\tempurl%
\url{https://doi.org/10.1145/3544549.3573802}
\showDOI{\tempurl}


\bibitem[Singh et~al\mbox{.}(2019)]%
        {Singh2019CameraInspiration}
\bibfield{author}{\bibinfo{person}{Dilpreet Singh}, \bibinfo{person}{Nina Rajcic}, \bibinfo{person}{Simon Colton}, {and} \bibinfo{person}{Jon McCormack}.} \bibinfo{year}{2019}\natexlab{}.
\newblock \showarticletitle{{Camera obscurer: Generative art for design inspiration}}.
\newblock \bibinfo{journal}{\emph{Lecture Notes in Computer Science (including subseries Lecture Notes in Artificial Intelligence and Lecture Notes in Bioinformatics)}}  \bibinfo{volume}{11453 LNCS} (\bibinfo{year}{2019}), \bibinfo{pages}{51--68}.
\newblock
\showISBNx{9783030166663}
\showISSN{16113349}
\urldef\tempurl%
\url{https://doi.org/10.1007/978-3-030-16667-0{\_}4/TABLES/3}
\showDOI{\tempurl}


\bibitem[Sio and Ormerod(2009)]%
        {Sio2009DoesReview}
\bibfield{author}{\bibinfo{person}{Ut~Na Sio} {and} \bibinfo{person}{Thomas~C. Ormerod}.} \bibinfo{year}{2009}\natexlab{}.
\newblock \showarticletitle{{Does Incubation Enhance Problem Solving? A Meta-Analytic Review}}.
\newblock \bibinfo{journal}{\emph{Psychological Bulletin}} \bibinfo{volume}{135}, \bibinfo{number}{1} (\bibinfo{date}{1} \bibinfo{year}{2009}), \bibinfo{pages}{94--120}.
\newblock
\showISSN{00332909}
\urldef\tempurl%
\url{https://doi.org/10.1037/A0014212}
\showDOI{\tempurl}


\bibitem[Smith et~al\mbox{.}(2013)]%
        {Smith2013ShiftingFixation}
\bibfield{author}{\bibinfo{person}{Melissa~A.B. Smith}, \bibinfo{person}{Robert~J. Youmans}, \bibinfo{person}{Brooke~G. Bellows}, {and} \bibinfo{person}{Matthew~S. Peterson}.} \bibinfo{year}{2013}\natexlab{}.
\newblock \showarticletitle{{Shifting the focus: An objective look at design fixation}}.
\newblock \bibinfo{journal}{\emph{Lecture Notes in Computer Science (including subseries Lecture Notes in Artificial Intelligence and Lecture Notes in Bioinformatics)}} \bibinfo{volume}{8012 LNCS}, \bibinfo{number}{PART 1} (\bibinfo{year}{2013}), \bibinfo{pages}{144--151}.
\newblock
\showISBNx{9783642392283}
\showISSN{03029743}
\urldef\tempurl%
\url{https://doi.org/10.1007/978-3-642-39229-0{\_}17/COVER}
\showDOI{\tempurl}


\bibitem[Smith and Linsey(2011)]%
        {Smith2011AFixation}
\bibfield{author}{\bibinfo{person}{Steven~M. Smith} {and} \bibinfo{person}{Julie Linsey}.} \bibinfo{year}{2011}\natexlab{}.
\newblock \showarticletitle{{A Three-Pronged Approach for Overcoming Design Fixation}}.
\newblock \bibinfo{journal}{\emph{The Journal of Creative Behavior}} \bibinfo{volume}{45}, \bibinfo{number}{2} (\bibinfo{date}{6} \bibinfo{year}{2011}), \bibinfo{pages}{83--91}.
\newblock
\showISSN{2162-6057}
\urldef\tempurl%
\url{https://doi.org/10.1002/J.2162-6057.2011.TB01087.X}
\showDOI{\tempurl}


\bibitem[Smith et~al\mbox{.}(1993)]%
        {Smith1993ConstrainingIn}
\bibfield{author}{\bibinfo{person}{Steven~M Smith}, \bibinfo{person}{Thomas~B Ward}, {and} \bibinfo{person}{Jays Schumacher}.} \bibinfo{year}{1993}\natexlab{}.
\newblock \showarticletitle{{Constraining effects of examples a creative generation task. In}}.
\newblock \bibinfo{journal}{\emph{Memorv {\&} Cognition}} \bibinfo{volume}{21}, \bibinfo{number}{6} (\bibinfo{year}{1993}), \bibinfo{pages}{837--845}.
\newblock


\bibitem[Terry and Hayfield(2021)]%
        {Terry2021EssentialsAnalysis}
\bibfield{author}{\bibinfo{person}{Gareth Terry} {and} \bibinfo{person}{Nikki Hayfield}.} \bibinfo{year}{2021}\natexlab{}.
\newblock \bibinfo{booktitle}{\emph{{Essentials of Thematic Analysis}}}.
\newblock \bibinfo{publisher}{American Psychological Association}, \bibinfo{address}{Washington, DC, USA}.
\newblock
\showISBNx{9781433835575}
\urldef\tempurl%
\url{https://uwe-repository.worktribe.com/output/7240960}
\showURL{%
\tempurl}


\bibitem[Vasconcelos et~al\mbox{.}(2018)]%
        {Vasconcelos2018IdeaExperiments}
\bibfield{author}{\bibinfo{person}{L.A. Vasconcelos}, \bibinfo{person}{M.A. Neroni}, \bibinfo{person}{C. Cardoso}, {and} \bibinfo{person}{N. Crilly}.} \bibinfo{year}{2018}\natexlab{}.
\newblock \showarticletitle{{Idea representation and elaboration in design inspiration and fixation experiments}}.
\newblock \bibinfo{journal}{\emph{International Journal of Design Creativity and Innovation}} \bibinfo{volume}{6}, \bibinfo{number}{1-2} (\bibinfo{year}{2018}), \bibinfo{pages}{93--113}.
\newblock
\urldef\tempurl%
\url{https://doi.org/10.1080/21650349.2017.1362360}
\showDOI{\tempurl}


\bibitem[Vasconcelos et~al\mbox{.}(2016)]%
        {Vasconcelos2016FluencyExplanation}
\bibfield{author}{\bibinfo{person}{L.A. Vasconcelos}, \bibinfo{person}{M.A. Neroni}, {and} \bibinfo{person}{N. Crilly}.} \bibinfo{year}{2016}\natexlab{}.
\newblock \showarticletitle{{Fluency results in design fixation experiments: An additional explanation}}. In \bibinfo{booktitle}{\emph{4th International Conference on Design Creativity, ICDC 2016}}. \bibinfo{publisher}{The Design Society}, \bibinfo{address}{Atlanta, GA, USA}, \bibinfo{pages}{1--8}.
\newblock
\showISBNx{9781904670827}


\bibitem[Vasconcelos et~al\mbox{.}(2017)]%
        {Vasconcelos2017InspirationGeneration}
\bibfield{author}{\bibinfo{person}{Luis~A Vasconcelos}, \bibinfo{person}{Carlos~C Cardoso}, \bibinfo{person}{Chih-Chun Chen}, {and} \bibinfo{person}{Nathan Crilly}.} \bibinfo{year}{2017}\natexlab{}.
\newblock \showarticletitle{{Inspiration and Fixation: The Influences of Example Designs and System Properties in Idea Generation}}.
\newblock \bibinfo{journal}{\emph{Journal of Mechanical Design}} \bibinfo{volume}{139}, \bibinfo{number}{3} (\bibinfo{year}{2017}), \bibinfo{pages}{031101}.
\newblock
\urldef\tempurl%
\url{https://doi.org/10.1115/1.4035540}
\showDOI{\tempurl}


\bibitem[Vasconcelos and Crilly(2016)]%
        {Vasconcelos2016InspirationChallenges}
\bibfield{author}{\bibinfo{person}{Luis~A. Vasconcelos} {and} \bibinfo{person}{Nathan Crilly}.} \bibinfo{year}{2016}\natexlab{}.
\newblock \showarticletitle{{Inspiration and fixation: Questions, methods, findings, and challenges}}.
\newblock \bibinfo{journal}{\emph{Design Studies}}  \bibinfo{volume}{42} (\bibinfo{date}{1} \bibinfo{year}{2016}), \bibinfo{pages}{1--32}.
\newblock
\showISSN{0142-694X}
\urldef\tempurl%
\url{https://doi.org/10.1016/J.DESTUD.2015.11.001}
\showDOI{\tempurl}


\bibitem[Vehtari et~al\mbox{.}(2021)]%
        {vehtari2021rhat}
\bibfield{author}{\bibinfo{person}{Aki Vehtari}, \bibinfo{person}{Andrew Gelman}, \bibinfo{person}{Daniel Simpson}, \bibinfo{person}{Bob Carpenter}, {and} \bibinfo{person}{Paul-Christian B{\"u}rkner}.} \bibinfo{year}{2021}\natexlab{}.
\newblock \showarticletitle{{Rank-Normalization, Folding, and Localization: An Improved $\widehat{R}$ for Assessing Convergence of MCMC (with Discussion)}}.
\newblock \bibinfo{journal}{\emph{Bayesian Analysis}} \bibinfo{volume}{16}, \bibinfo{number}{2} (\bibinfo{year}{2021}), \bibinfo{pages}{667 -- 718}.
\newblock
\urldef\tempurl%
\url{https://doi.org/10.1214/20-BA1221}
\showDOI{\tempurl}


\bibitem[Verheijden and Funk(2023)]%
        {Verheijden2023CollaborativeAI}
\bibfield{author}{\bibinfo{person}{Mathias~Peter Verheijden} {and} \bibinfo{person}{Mathias Funk}.} \bibinfo{year}{2023}\natexlab{}.
\newblock \showarticletitle{{Collaborative Diffusion: Boosting Designerly Co-Creation with Generative AI}}. In \bibinfo{booktitle}{\emph{Conference on Human Factors in Computing Systems - Proceedings}}. \bibinfo{publisher}{Association for Computing Machinery}, \bibinfo{address}{New York, NY, USA}, \bibinfo{pages}{1--8}.
\newblock
\showISBNx{9781450394222}
\urldef\tempurl%
\url{https://doi.org/10.1145/3544549.3585680}
\showDOI{\tempurl}


\bibitem[Viswanathan and Linsey(2012)]%
        {Viswanathan2012DesignCost}
\bibfield{author}{\bibinfo{person}{Vimal Viswanathan} {and} \bibinfo{person}{Julie Linsey}.} \bibinfo{year}{2012}\natexlab{}.
\newblock \showarticletitle{{Design Fixation in Physical Modeling: An Investigation on the Role of Sunk Cost}}.
\newblock \bibinfo{journal}{\emph{Proceedings of the ASME Design Engineering Technical Conference}}  \bibinfo{volume}{9} (\bibinfo{date}{6} \bibinfo{year}{2012}), \bibinfo{pages}{119--130}.
\newblock
\showISBNx{9780791854860}
\urldef\tempurl%
\url{https://doi.org/10.1115/DETC2011-47862}
\showDOI{\tempurl}


\bibitem[Viswanathan et~al\mbox{.}(2016)]%
        {Viswanathan2016AFixation}
\bibfield{author}{\bibinfo{person}{V. Viswanathan}, \bibinfo{person}{M. Tomko}, {and} \bibinfo{person}{J. Linsey}.} \bibinfo{year}{2016}\natexlab{}.
\newblock \showarticletitle{{A study on the effects of example familiarity and modality on design fixation}}.
\newblock \bibinfo{journal}{\emph{Artificial Intelligence for Engineering Design, Analysis and Manufacturing: AIEDAM}} \bibinfo{volume}{30}, \bibinfo{number}{2} (\bibinfo{year}{2016}), \bibinfo{pages}{171--184}.
\newblock
\urldef\tempurl%
\url{https://doi.org/10.1017/S0890060416000056}
\showDOI{\tempurl}


\bibitem[Wagenmakers et~al\mbox{.}(2011)]%
        {wagenmakers2011psychologists}
\bibfield{author}{\bibinfo{person}{Eric-Jan Wagenmakers}, \bibinfo{person}{Ruud Wetzels}, \bibinfo{person}{Denny Borsboom}, {and} \bibinfo{person}{Han~LJ Van Der~Maas}.} \bibinfo{year}{2011}\natexlab{}.
\newblock \showarticletitle{Why psychologists must change the way they analyze their data: the case of psi: comment on Bem (2011).}
\newblock  (\bibinfo{year}{2011}).
\newblock


\bibitem[Ward(1994)]%
        {Ward1994StructuredGeneration}
\bibfield{author}{\bibinfo{person}{T.~B. Ward}.} \bibinfo{year}{1994}\natexlab{}.
\newblock \showarticletitle{{Structured Imagination: the Role of Category Structure in Exemplar Generation}}.
\newblock \bibinfo{journal}{\emph{Cognitive Psychology}} \bibinfo{volume}{27}, \bibinfo{number}{1} (\bibinfo{date}{8} \bibinfo{year}{1994}), \bibinfo{pages}{1--40}.
\newblock
\showISSN{0010-0285}
\urldef\tempurl%
\url{https://doi.org/10.1006/COGP.1994.1010}
\showDOI{\tempurl}


\bibitem[Williford et~al\mbox{.}(2023)]%
        {Williford2023ExploringIdeation}
\bibfield{author}{\bibinfo{person}{Blake Williford}, \bibinfo{person}{Samantha Ray}, \bibinfo{person}{Jung~In Koh}, \bibinfo{person}{Josh Cherian}, \bibinfo{person}{Paul Taele}, {and} \bibinfo{person}{Tracy Hammond}.} \bibinfo{year}{2023}\natexlab{}.
\newblock \showarticletitle{{Exploring Creativity Support for Concept Art Ideation}}. In \bibinfo{booktitle}{\emph{Extended Abstracts of the 2023 CHI Conference on Human Factors in Computing Systems}}. \bibinfo{publisher}{Association for Computing Machinery}, \bibinfo{address}{New York, NY, USA}, \bibinfo{pages}{1--7}.
\newblock
\showISBNx{9781450394222}
\urldef\tempurl%
\url{https://doi.org/10.1145/3544549.3585684}
\showDOI{\tempurl}


\bibitem[Wingstr{\"{o}}m et~al\mbox{.}(2022)]%
        {Wingstrom2022RedefiningArtists}
\bibfield{author}{\bibinfo{person}{Roosa Wingstr{\"{o}}m}, \bibinfo{person}{Johanna Hautala}, {and} \bibinfo{person}{Riina Lundman}.} \bibinfo{year}{2022}\natexlab{}.
\newblock \showarticletitle{{Redefining Creativity in the Era of AI? Perspectives of Computer Scientists and New Media Artists}}.
\newblock \bibinfo{journal}{\emph{Creativity Research Journal}} (\bibinfo{year}{2022}), \bibinfo{pages}{1--17}.
\newblock
\showISSN{10400419}
\urldef\tempurl%
\url{https://doi.org/10.1080/10400419.2022.2107850}
\showDOI{\tempurl}


\bibitem[Youmans(2011a)]%
        {Youmans2011DesignFixation}
\bibfield{author}{\bibinfo{person}{Robert~J. Youmans}.} \bibinfo{year}{2011}\natexlab{a}.
\newblock \showarticletitle{{Design Fixation in the Wild: Design Environments and Their Influence on Fixation}}.
\newblock \bibinfo{journal}{\emph{The Journal of Creative Behavior}} \bibinfo{volume}{45}, \bibinfo{number}{2} (\bibinfo{date}{6} \bibinfo{year}{2011}), \bibinfo{pages}{101--107}.
\newblock
\showISSN{2162-6057}
\urldef\tempurl%
\url{https://doi.org/10.1002/J.2162-6057.2011.TB01089.X}
\showDOI{\tempurl}


\bibitem[Youmans(2011b)]%
        {Youmans2011TheFixation}
\bibfield{author}{\bibinfo{person}{Robert~J. Youmans}.} \bibinfo{year}{2011}\natexlab{b}.
\newblock \showarticletitle{{The effects of physical prototyping and group work on the reduction of design fixation}}.
\newblock \bibinfo{journal}{\emph{Design Studies}} \bibinfo{volume}{32}, \bibinfo{number}{2} (\bibinfo{date}{3} \bibinfo{year}{2011}), \bibinfo{pages}{115--138}.
\newblock
\showISSN{0142-694X}
\urldef\tempurl%
\url{https://doi.org/10.1016/J.DESTUD.2010.08.001}
\showDOI{\tempurl}


\bibitem[Youmans and Arciszewski(2014a)]%
        {Youmans2014DesignColors}
\bibfield{author}{\bibinfo{person}{Robert~J. Youmans} {and} \bibinfo{person}{Tomasz Arciszewski}.} \bibinfo{year}{2014}\natexlab{a}.
\newblock \showarticletitle{{Design Fixation: A Cloak of Many Colors}}. In \bibinfo{booktitle}{\emph{Design Computing and Cognition '12}}. \bibinfo{publisher}{Springer}, \bibinfo{address}{Dordrecht}, \bibinfo{pages}{115--129}.
\newblock
\urldef\tempurl%
\url{https://doi.org/10.1007/978-94-017-9112-0}
\showDOI{\tempurl}


\bibitem[Youmans and Arciszewski(2014b)]%
        {Youmans2014DesignPrevention}
\bibfield{author}{\bibinfo{person}{Robert~J. Youmans} {and} \bibinfo{person}{Thomaz Arciszewski}.} \bibinfo{year}{2014}\natexlab{b}.
\newblock \showarticletitle{{Design fixation: Classifications and modern methods of prevention}}.
\newblock \bibinfo{journal}{\emph{AI EDAM}} \bibinfo{volume}{28}, \bibinfo{number}{2} (\bibinfo{year}{2014}), \bibinfo{pages}{129--137}.
\newblock
\showISSN{0890-0604}
\urldef\tempurl%
\url{https://doi.org/10.1017/S0890060414000043}
\showDOI{\tempurl}


\end{thebibliography}

\appendix 

\section*{APPENDIX}

\setcounter{table}{0}
\renewcommand{\thetable}{A.\arabic{table}}
	
\section{The Mean Scores and Standard Errors of the NASA Task Load Index (NASA-TLX) Scales}

\begin{table}[htbp]

    \centering
     \caption{\hl{The mean scores and standard error for each NASA TLX scale (mental demand, physical demand, temporal demand, performance demand, effort demand, frustration demand) in the three conditions: No support, Image Search and GenAI}}
    \begin{tabular}{llll}
    \toprule
 & \multicolumn{3}{c}{\textbf{Mean scores \& Std. Error}}\\
 
         \textbf{NASA TLX}&  \textbf{No Support}&  \textbf{Image Search}& \textbf{GenAI}\\
         \midrule
         Mental Demand
&  4.1 (0.3) &  4.0 (0.4) & 4.2 (0.4)\\
         Physical Demand
&  2.2 (0.3)&  2.6 (0.3)& 2.6 (0.3)\\
         Temporal Demand
&  4.3 (0.4) &  3.7 (0.4)& 4.4 (0.4)\\
         Performance
&  4.1 (0.3) &  5.2 (0.3) & 4.7 (0.3)\\
         Effort
&  4.5 (0.3) &  4.0 (0.4) & 4.0 (0.2)\\
         Frustration
&  3.1 (0.4)&  2.0 (0.3) & 2.2 (0.3)\\
\bottomrule
    \end{tabular}
    \label{tab:NASA_TLX}

\end{table}

\end{document}